\documentclass{aa}  
\usepackage{graphicx}
\usepackage{txfonts}
\usepackage{amsmath}
\usepackage{booktabs} 
\usepackage{dcolumn}  
\usepackage{fixltx2e} 
\usepackage{color}
\usepackage[UKenglish]{isodate}

\cleanlookdateon 
\newcolumntype{d}[1]{D{.}{.}{#1}} 
\def\myeol{\\}
\providecommand{\gaia}{Gaia}
\providecommand{\gdr}[1]{Gaia\,DR{#1}}
\providecommand{\gmag}{\ensuremath{G}}
\providecommand{\teff}{\ensuremath{T_{\rm eff}}}
\providecommand{\nenc}{\ensuremath{n_{\rm enc}}}
\providecommand{\ncor}{\ensuremath{n_{\rm cor}}}

\def\tph{t_{\rm ph}}
\def\dph{d_{\rm ph}}
\def\vph{\mathrm v_{\rm ph}}

\def\tphmed{t_{\rm ph}^{\rm med}}
\def\dphmed{d_{\rm ph}^{\rm med}}
\def\vphmed{\mathrm v_{\rm ph}^{\rm med}}

\def\ra{\alpha}
\def\dec{\delta}
\providecommand{\parallax}{\varpi}
\def\pmra{\mu_{\ra\ast}}
\def\pmdec{\mu_\dec}
\def\propm{\mu}

\def\vr{\mathrm v_r}
\def\sigvr{\sigma(\vr)}

\def\mass{M}

\def\fmod{F_{\rm mod}}
\def\fexp{F_{\rm exp}}

\def\mas{mas}
\def\kms{km\,s$^{-1}$}

\def\maspyr{\mas\,yr$^{-1}$}
\def\Msol{M$_\odot$}

\begin{document}

\defcitealias{2015A&A...575A..35B}{paper~1} 
\defcitealias{2018A&A...609A...8B}{paper~2}


\title{New stellar encounters discovered in the second Gaia data release}
\titlerunning{Close encounters to the Sun in Gaia-DR2}
\author{C.A.L.~Bailer-Jones, J.\ Rybizki, R.\ Andrae, M.\ Fouesneau}
\authorrunning{Bailer-Jones et al.}
\institute{Max Planck Institute for Astronomy, K\"onigstuhl 17, 69117 Heidelberg, Germany}
\date{Submitted 19 May 2018. Revised 17 June 2018. Accepted 19 June 2018.}
\abstract{ 
Passing stars may play an important role in the evolution of our solar system.
We search for close stellar encounters to the Sun among all 7.2 million stars in \gdr{2} that have six-dimensional phase space data. We characterize encounters by integrating their orbits through a Galactic potential and propagating the correlated uncertainties via a Monte Carlo resampling. After filtering to remove spurious data, we find 694 stars that have median (over uncertainties) closest encounter distances within 5\,pc, all occurring within 15\,Myr from now. 26 of these have at least a 50\% chance of coming closer than 1\,pc (and 7 within 0.5\,pc), all but one of which are newly discovered here.
We confirm some and refute several other previously-identified encounters, confirming suspicions about their data.
The closest encounter in the sample is \object{Gl 710}, which has a 95\% probability of coming closer than 0.08\,pc (17\,000\,AU).
Taking mass estimates obtained from \gaia\ astrometry and multiband photometry for essentially all encounters, we find that \object{Gl 710} also has the largest impulse on the Oort cloud.
Using a Galaxy model, we compute the completeness of the \gdr{2} encountering sample as a function of perihelion time and distance. 
Only 15\% of encounters within 5\,pc occurring within $\pm 5$\,Myr of now
have been identified, mostly due to the lack of radial velocities for faint and/or cool stars. Accounting for the incompleteness, we infer the present rate of encounters within 1\,pc to be $19.7 \pm 2.2$ per Myr, a quantity expected to scale quadratically with the encounter distance out to at least several pc.  
Spuriously large parallaxes in our sample from imperfect filtering would tend to inflate both
the number of encounters found and this inferred rate. The magnitude of this effect is hard to quantify.
}
\keywords{Oort cloud -- methods: analytical, statistical  -- solar neighbourhood -- stars: kinematics and dynamics -- surveys: Gaia} 
\maketitle

\section{Introduction}

The first convincing evidence for relative stellar motion came from Edmund Halley in 1718.
Since then -- if not well before -- people have wondered how close other stars may come to our own. Several studies over the past quarter century have used proper motions and radial velocities to answer this question
\citep[][]{1994QJRAS..35....1M, 1996EM&P...72...19M, 1999AJ....117.1042G, 2001A&A...379..634G, 2006A&A...449.1233D, 2010AstL...36..220B, 2010AstL...36..816B, 2011MNRAS.418.1272J, 2015A&A...575A..35B, 2015MNRAS.449.2459D, 2015ApJ...800L..17M, 2016A&A...595L..10B, 2017AstL...43..559B, 2017ARep...61..883B, 2018A&A...609A...8B}.

Other than being interesting in their own right, close encounters may have played a significant role in the evolution of our solar system, in particular of the Oort cloud. This may also have had implications for the development of life, since a strong perturbation of the Oort cloud by an encountering star could push comets into the inner solar system. An ensuing collision with the Earth could be catastrophic enough to cause a mass extinction. Such a fate probably befell the dinosaurs 65\,Myr ago.  Studies of stellar encounters and comet impacts on the Earth can also be used to learn about the general hazards for life on exoplanets.

The recent publication of the second \gaia\ data release (\gdr{2}; \citealt{2018arXiv180409365G}) is a boon to encounter studies.  It contains six-dimensional (6D) kinematic data -- position, parallax, proper motion, and radial velocity -- for 7.2 million stars. This is 22 times larger than our previous encounter study using TGAS in the first \gaia\ data release.
Furthermore, TGAS contained no radial velocities, so these had to be obtained by cross matching to external catalogues.

Here we report on results of looking for close encounters in \gdr{2}. Our approach follows very closely that taken by
\cite{2015A&A...575A..35B} \citepalias{2015A&A...575A..35B}
 for Hipparcos and \cite{2018A&A...609A...8B} 
\citepalias{2018A&A...609A...8B} for TGAS. The main differences here are:
\begin{itemize}
\item We use only \gaia\ data. This avoids the complications of having to obtain radial velocities from several different catalogues, which resulted in some data heterogeneity and duplicate sources. 
\item Using mass estimates from multiband photometry and \gdr{2} astrometry in Fouesneau et al.\ (in preparation), we estimate the momentum transfer from an encountering star to Oort cloud comets for essentially every encounter.
\item We account for the incompleteness of our encounter sample by sampling from a self-consistent spatial and kinematic Galaxy model. This is more realistic that the analytic model developed in \citetalias{2018A&A...609A...8B}.
\end{itemize}
The main limitation in the number of encounter candidates in this study is still the availability of radial velocities. While \gdr{2} contains five-parameter astrometry for 1.33 billion stars, only 7.2 million have published radial velocities. These are predominantly
brighter than $\gmag=14$\,mag, and are also limited to the approximate \teff\ range of $3550 < \teff / {\rm K} < 6900$
\citep{2018arXiv180409372K}.
(This is simply the temperature range of the templates used. \teff\ estimates in \gdr{2} are described in \citealt{2018arXiv180409374A}).

In section~\ref{sec:procedure} we describe how we select our sample and
infer the distribution of the encounter parameters for each candidate. We analyse the results in section~\ref{sec:results}, discussing new (and dubious) cases, and highlighting disagreement with earlier results. 
Issues of spurious data and imperfect filtering we discuss in subsections~\ref{sec:filtering} and~\ref{sec:cmd}.
In section~\ref{sec:completeness} we introduce the new completeness model and use this to derive the completeness-corrected encounter rate. We conclude in section~\ref{sec:conclusions} with a brief discussion.

\section{Identification and characterization of close encounters}\label{sec:procedure}

\subsection{Initial selection and orbital integration}

We searched the \gaia\ archive for all stars which would approach within 10\,pc of the Sun under the assumption that they move on unaccelerated paths relative to the solar system (the so-called ``linear motion approximation'', LMA, defined in \citetalias{2015A&A...575A..35B}). The archive ADQL query for this is in appendix~\ref{appendix:query}. This yielded 3865 encounter candidates, which we call the {\em unfiltered sample}. 
In both this selection and the orbital integration used later, we estimate distance using inverse parallax rather than doing a proper inference \citep{2015PASP..127..994B, 2018arXiv180410121B}. This is acceptable here, because for the filtered sample we define later, 95\% have fractional parallax uncertainties below 0.08 (99\% below 0.14; the largest is 0.35), meaning the simple inversion is a reasonably good approximation.

The \gaia\ data are used as they are, other than we added 0.029\,mas to the parallaxes to accommodate the global parallax zeropoint \cite{2018arXiv180409366L}. (Neglecting this would remove 83 sources from the unfiltered sample.)
There is no evidence for a proper motion zeropoint offset \citep{2018arXiv180409375A}. While there may be a systematic difference between the \gaia\ radial velocities and other catalogues of up to 0.5\,\kms, the origin of this is unclear \citep{2018arXiv180409372K}. This is of the order of the gravitational redshifts which are also not corrected for. None of the \gaia\ uncertainties have been adjusted to accommodate a possible over- or underestimate of their values (e.g.\ missing RMS systematics).

To get more precise encounter parameters for the set of candidates, we integrated the orbits of the unfiltered sample through a smooth Galactic potential forward and backwards in time. (It was shown in \citetalias{2015A&A...575A..35B} that the deviation of an orbit due to perturbations by individual stars can be neglected.)
We use the same procedure and model as described in \citetalias{2018A&A...609A...8B}. 
The Galactic potential, described in detail in \citetalias{2015A&A...575A..35B}, is a three-component axisymmetric model. The bar and spiral arms are not included, partly because their properties are not well determined, but mostly because using the same potential as in our previous studies eases comparison of results. As the orbit segments up to encounter are generally short compared to the scale lengths in the model, the exact choice of potential will have only a small impact on the encounter parameters for most stars.

In order to accommodate and propagate the uncertainties in the data, we draw 2000 samples from the 6D covariant probability density function (PDF) over the data -- position, parallax, proper motion, and radial velocity -- for each star and integrate the orbits of each of these ``surrogates'' through the potential. The distribution of the perihelion time, distance, and speed over the surrogates for each star is used to characterize the encounter (in the next section). Comparisons of encounter parameters computed with the LMA, orbital integration of the nominal data, and orbital integration of the surrogates were shown in papers 1 and 2. Due to the nonlinear transformation from astrometric measurements to perihelion parameters, neglecting the full PDF can lead to erroneous results (and not just erroneous uncertainties), in particular for stars with long travel times.

\subsection{Filtering on astrometric solution quality metrics}\label{sec:filtering}

There are many reasons why astrometric solutions in \gdr{2} may be ``wrong'' for some stars, in the sense that the reported uncertainties may not be representative of the true uncertainties.  The main reasons are: neglect of accelerated motions (i.e.\ unseen companions); cross-matching errors leading to the inclusion of observations of other sources (spurious data); a poor correction of the so-called ``DOF bug'' (see appendix A of \citealt{2018arXiv180409366L}).  These can lead to erroneous estimates of the quantities or their uncertainties.

Various metrics on the astrometric solution are reported in \gdr{2} to help identify good solutions.  \cite{2018arXiv180409366L} discuss some of these and give an example of a set of cuts which may be used to define a conservative sample, i.e.\ an agressive removal of poor solutions (e.g.\ in their Figure C.2).  This is not appropriate for our work, however, because we are looking to determine the encounter rate, not just find the most reliable encounters.  While quantiles on the various metrics are easily measured, there is no good model for the expected distribution of these for only non-spurious results.
Concepts like ``the reduced $\chi^2$ should be about one'' are simplistic at best (and statistically questionable), and also 
don't tell us whether deviations from this are ``wrong'' or just have mildly deviant astrometry or slightly underestimated uncertainties.
Even a highly significant astrometric excess noise of a few mas may be of little consequence if the parallax and proper motion are large.
It is therefore difficult to make a reliable cut. The main quality metric of interest here is the ``unit weight error'', defined in appendix A of \cite{2018arXiv180409366L} as $u=(\chi^2/\nu)^{1/2}$, where $\chi^2$ is the metric {\tt astrometric\_chi2\_al}
and $\nu$ is the degrees of freedom, equal to {\tt astrometric\_n\_good\_obs\_al}$-5$. For our unfiltered sample, $u$ correlates quite strongly with the {\tt astrometric\_excess\_noise} and reasonably well, but less tightly, with {\tt astrometric\_excess\_noise\_sig}. There is no correlation between $u$ and {\tt visibility\_periods\_used}. 
Figure C.2 of \cite{2018arXiv180409366L} plots $u$ against \gmag. 
We see $u$ increasing for brighter sources, albeit it with a lot of scatter.
The full range of $u$ in our unfiltered sample is 0.63 to 122, and 
the brighest star in our unfiltered sample (\object{\gdr{2} 5698015743040715264} = \object{rho Puppis}) has $\gmag=2.67$\,mag. This corresponds to $u=35$ for the selection in Figure C.2, so we only retain sources with $u<35$. 
This moderately liberal cut reduces the sample size to 3465.
\object{Rho Puppis} itself has $u=29$,  {\tt astrometric\_excess\_noise}\,=\,2.4\,\mas, and
 {\tt astrometric\_excess\_noise\_sig}\,=\,2458, yet its parallax and proper motion in \gdr{2} agree with those from Hipparcos-2 \citep{2007ASSL..350.....V} to within about 2\%. A more agressive cut would clearly be removing genuine solutions.
To ensure that the astrometric solutions are reasonably overdetermined, we further require that 
{\tt visibility\_periods\_used} is at least 8. According to \cite{2018arXiv180409375A},
this should help to remove the most spurious proper motions. 
This cuts the sample down to 3379.

Large uncertainties in the data, provided they are representative of the true uncertainty, are not a problem per se because they are accommodated by the resampling we used to map the PDF of the perihelion parameters. 
We will also accommodate this PDF when we compute the completeness and encounter rate in section \ref{sec:completeness}.
We therefore do not filter out sources due to large uncertainties. 

One should be very careful about filtering out extreme values of the data
just because they are extreme. \gdr{2} is known to include sources with implausibly large parallaxes. For example, there are 21 sources with $\parallax>1000$\,mas (which would place them all nearer than Proxima Centauri). However, as they all have $\gmag>19$\,mag none of them appear in any of our selections.
Other, less extreme, spurious values are impossible to identify without using additional information.
Spuriously large proper motions are less of a problem, because close encounter stars generally have relatively small proper motions (unless they are currently very close to encounter).
In contrast, stars which come close tend to be those that currently have large radial velocities (compared to their transverse velocities).
The radial velocity processing for \gdr{2} may have produced spuriously large radial velocities (the pipeline can produce values up to $\pm$1000\,\kms).\cite{2018arXiv180409372K} report they were hard to verify due to the absence of observations of standards with  radial velocities larger than 550\,kms. They further state that the precisions are lower for stars with  $|\vr|>175$\,\kms, but not that the radial velocities themselves are problematic. Moreover, they visually inspected all results with  $|\vr|>500$\,\kms\ and removed suspicious results.
At this point in the filtering we have
only seven stars with $|\vr|>500$\,\kms, six of which were determined by just two focal plane transits ({\tt rv\_nb\_transits}\,=\,2; the minimum for inclusion in \gdr{2}) whereas the median number for this sample is 8.
They do not have particularly large uncertainties, because \gdr{2} only includes stars for which $|\sigvr| < 20$\,\kms.
It is tempting to filter out stars with a small number of transits, but this also removes valid measurements, such as that for the closest known encounter, \object{Gl 710}, which has two transits and a radial velocity consistent with non-\gaia\ measurements.
We do not filter on {\tt rv\_nb\_transits}. 

We define the remaning set of 3379 stars as the {\em filtered sample}. It comprises all stars that approach within 10\,pc of the Sun according to the LMA (but not necessarily the orbital integration) and which satisfy the filters $u < 35$ and {\tt visibility\_periods\_used}\,$\geq 8$. 
When it comes to considering the completeness-corrected encounter rate (section~\ref{sec:encounter_frequency}), we shall further limit this sample to stars with $\gmag<12.5$\,mag, which contains 2522 stars.

\section{Close encounters found in \gdr{2}}\label{sec:results}

We first look at the overall results, then discuss individual stars found approaching within 1\,pc, and finally those close encounters from papers 1 and 2 not found in the present study.

\subsection{Overall results}

\begin{figure}
\begin{center}
\includegraphics[width=0.5\textwidth, angle=0]{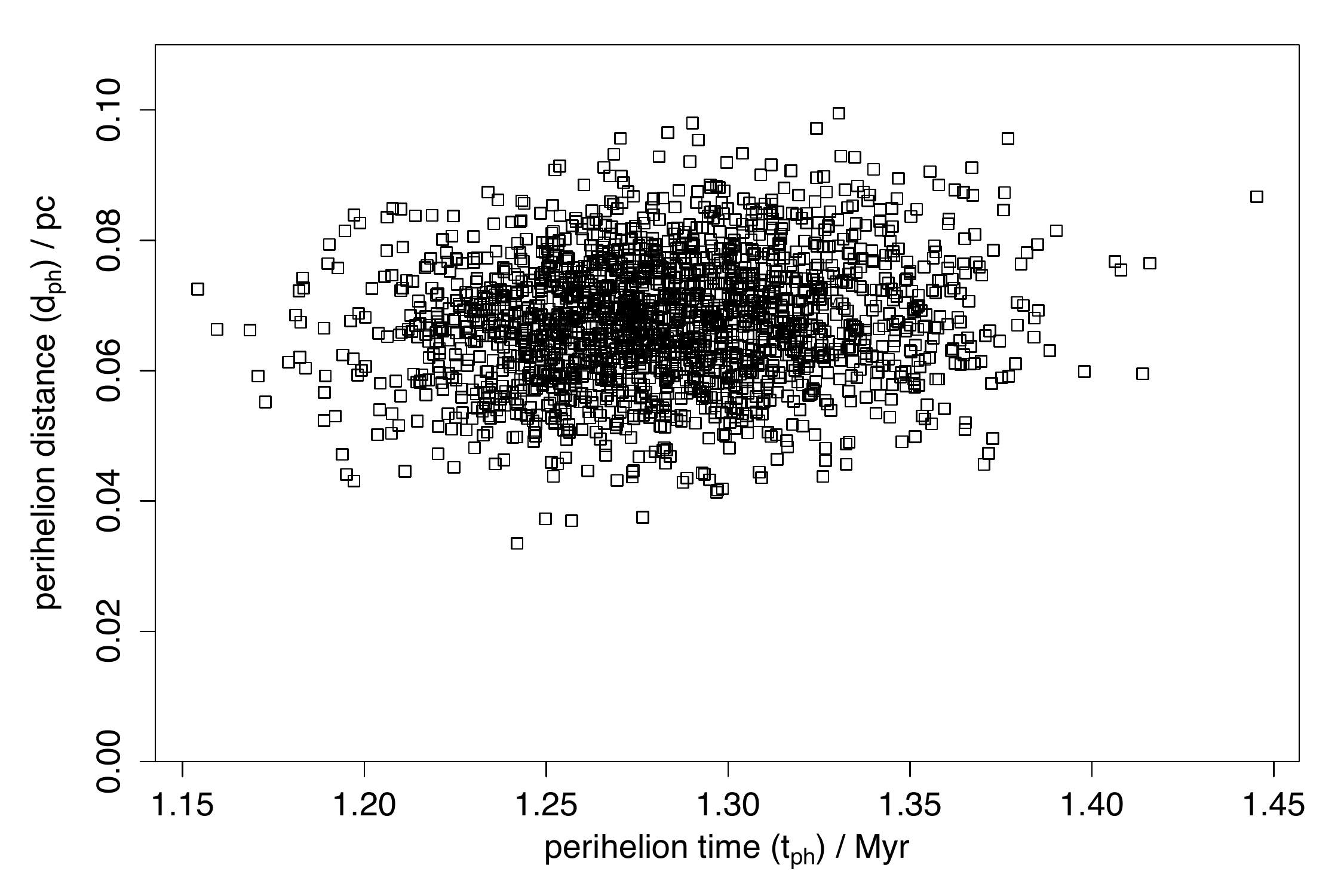}
\caption{The distribution of the perihelion parameters for the 2000 surrogates used to characterize the encounter of 
\object{Gl 710} = \object{\gdr{2} 4270814637616488064}.
\label{fig:dph_vs_tph_surrogates_Gl710}}
\end{center}
\end{figure} 

The perihelion for each star is described by the distribution of the 2000 surrogates in the perihelion time, $\tph$, distance, $\dph$, and speed $\vph$. 
The distribution for one particular star is shown in Figure~\ref{fig:dph_vs_tph_surrogates_Gl710}.
As in papers 1 and 2 we summarize these using the median, and characterize their uncertainty using the 5th and 95th percentiles (which together form a 90\% confidence interval, CI).  The number of stars coming within various perihelion distances is shown in Table~\ref{tab:dphSummary} (see also Figure~\ref{fig:fobs_dph} later).  Summary perihelion data for those stars with $\dphmed < 1$\,pc are shown in Table~\ref{tab:periStats} (the online table at CDS reports higher numerical precisions for the whole filtered sample). Negative times indicate past encounters. The magnitude, colour, and quality metrics from \gdr{2} are shown in Table~\ref{tab:extras}. Some of these encounters we have cause to disregard (discussed later).

\begin{table}
\begin{center}
\caption{The number of stars in the filtered sample found by the orbit integration to have $\dphmed < \dph^{\rm max}$ (for any $\tphmed$).
Stars with potentially problematic data have not been excluded.
\label{tab:dphSummary}}
\begin{tabular}{ d{1} r }
\toprule
\multicolumn{1}{c}{$\dph^{\rm max}$\,/\,pc}  & No.\ stars \\
\midrule
{\rm \infty} & 3379 \\
10 & 2548 \\
5 & 694 \\
3 & 283 \\
2 & 129 \\
1 & 31 \\
0.5 & 8 \\
0.25 & 3 \\
\bottomrule
\end{tabular}
\end{center}
\end{table}

\begin{table*}
\centering
\tiny{
\caption{Perihelion parameters for all stars with a median perihelion distance (median of the samples; $\dphmed$) below 1\,pc, sorted by this value.
The first column is the \gdr{2} source ID. Columns 2, 5, and 8 are
$\tphmed$, $\dphmed$, and $\vphmed$ respectively. The columns labelled 5\% and 95\% are the bounds of corresponding confidence intervals.
Columns 11--16 list the parallax ($\parallax$, plus the 0.029\,mas zeropoint offset), total proper motion ($\propm$), and radial velocity ($\vr$) along with their 1-sigma uncertainties. 
Column 17 is the estimated mass of the star from Fouesneau et al.\ (in preparation); NA indicates it is missing. The formal 1-sigma uncertainties in the masses are a few percent (systematics are likely to be higher).
Those encounter results we consider bogus are marked with the dagger symbol in the final column.
These are discussed in the text. Other may also be dubious: quality metrics from \gdr{2} can be found in Table~\ref{tab:extras}.
The online table at CDS includes all 3379 stars in the filtered sample and reports some columns to a higher numerical precision.
\label{tab:periStats}}
\tabcolsep=0.14cm
\begin{tabular}{*{18}{r}}
\toprule
1 & 2 & 3 & 4 & 5 & 6 & 7 & 8 & 9 & 10 & 11 & 12 & 13 & 14 & 15 & 16 & 17 & \\
\midrule
\gdr{2} source ID & \multicolumn{3}{c}{$\tph$ / kyr} &  \multicolumn{3}{c}{$\dph$ / pc} &  \multicolumn{3}{c}{$\vph$ / \kms} &
 $\parallax$ & $\sigma(\parallax)$ & $\propm$ & $\sigma(\propm)$ & $\vr$ & $\sigma(\vr)$ & $\mass$ & \\
         & med & 5\% & 95\% & med & 5\% & 95\% & med & 5\% & 95\% & \multicolumn{2}{c}{\mas} & \multicolumn{2}{c}{\maspyr} &  \multicolumn{2}{c}{\kms} & \Msol & \\
\midrule
 4270814637616488064 &     1281 &     1221 &     1352 &   0.068 &   0.052 &   0.084 &    14.5 &    13.8 &    15.2 &    52.55 &     0.05 &     0.46 &     0.11 &   -14.5 &     0.4 &   0.68 &  \myeol 
  955098506408767360 &     -737 &     -817 &     -672 &   0.151 &   0.043 &   0.279 &    38.5 &    34.7 &    42.1 &    34.53 &     0.61 &     0.82 &     1.21 &    38.5 &     2.1 &   1.26 &  \myeol 
 5571232118090082816 &    -1162 &    -1173 &    -1151 &   0.232 &   0.199 &   0.264 &    82.3 &    81.5 &    83.1 &    10.23 &     0.02 &     0.42 &     0.05 &    82.2 &     0.5 &   0.82 &  \myeol 
 2946037094755244800 &     -908 &    -1054 &     -795 &   0.338 &   0.101 &   0.662 &    42.1 &    36.9 &    47.2 &    25.66 &     1.12 &     1.34 &     1.82 &    42.1 &     3.2 &     NA &  \myeol 
 4071528700531704704 &      438 &      268 &     1109 &   0.374 &   0.213 &   1.037 &    44.2 &    16.9 &    71.7 &    50.43 &     0.89 &     8.91 &     1.92 &   -44.5 &    17.1 &   1.00 &  \myeol 
  510911618569239040 &    -2788 &    -2853 &    -2731 &   0.429 &   0.368 &   0.494 &    26.5 &    25.9 &    27.1 &    13.23 &     0.04 &     0.56 &     0.05 &    26.4 &     0.3 &   1.07 &  \myeol 
  154460050601558656 &      372 &      332 &      418 &   0.444 &   0.225 &   0.734 &   233.5 &   218.2 &   247.8 &    11.29 &     0.67 &     2.16 &     0.99 &  -233.1 &     8.9 &     NA &  $\dagger$ \myeol 
 6608946489396474752 &    -2757 &    -2826 &    -2694 &   0.491 &   0.301 &   0.681 &    45.3 &    44.3 &    46.3 &     7.89 &     0.05 &     0.67 &     0.12 &    44.2 &     0.6 &   0.82 &  \myeol 
 3376241909848155520 &     -450 &     -516 &     -396 &   0.508 &   0.276 &   0.773 &    79.9 &    70.8 &    88.9 &    27.18 &     1.09 &     5.96 &     2.56 &    79.9 &     5.6 &   1.04 &  \myeol 
 1791617849154434688 &    -1509 &    -1533 &    -1487 &   0.579 &   0.503 &   0.654 &    56.4 &    55.6 &    57.2 &    11.49 &     0.04 &     0.88 &     0.09 &    56.3 &     0.5 &   0.80 &  \myeol 
 4265426029901799552 &     -655 &     -686 &     -627 &   0.580 &   0.368 &   0.791 &    46.6 &    46.3 &    46.9 &    32.05 &     0.88 &     5.93 &     1.95 &    46.6 &     0.2 &   0.49 &  \myeol 
 5261593808165974784 &     -897 &     -916 &     -879 &   0.636 &   0.608 &   0.664 &    71.1 &    69.7 &    72.6 &    15.32 &     0.02 &     2.33 &     0.06 &    71.0 &     0.9 &   0.55 &  \myeol 
 5896469620419457536 &     6680 &     5880 &     7550 &   0.657 &   0.250 &   1.088 &    16.8 &    14.9 &    19.1 &     8.69 &     0.03 &     0.75 &     0.08 &   -16.9 &     1.3 &   0.62 &  \myeol 
 4252068750338781824 &      886 &      452 &     3177 &   0.668 &   0.352 &   3.257 &    27.6 &     5.8 &    51.7 &    38.86 &     0.61 &     5.74 &     1.24 &   -27.7 &    14.2 &   0.89 &  \myeol 
 1949388868571283200 &     -710 &     -738 &     -685 &   0.673 &   0.591 &   0.762 &   347.4 &   336.5 &   357.9 &     3.96 &     0.04 &     0.80 &     0.05 &   347.3 &     6.5 &     NA &  $\dagger$ \myeol 
 1802650932953918976 &     5341 &     4770 &     6033 &   0.740 &   0.213 &   1.370 &    53.0 &    47.1 &    59.2 &     3.46 &     0.04 &     0.76 &     0.07 &   -52.6 &     3.9 &   0.98 &  \myeol 
 3105694081553243008 &     -715 &     -790 &     -656 &   0.760 &   0.516 &   0.990 &    38.3 &    35.1 &    41.4 &    35.72 &     0.97 &     7.44 &     2.01 &    38.4 &     1.9 &   0.75 &  \myeol 
 5231593594752514304 &       89 &       89 &       89 &   0.815 &   0.807 &   0.822 &   715.9 &   714.1 &   717.6 &    15.35 &     0.03 &    29.87 &     0.07 &  -715.8 &     1.0 &   0.67 &  $\dagger$ \myeol 
 4472507190884080000 &     1836 &     1210 &     3382 &   0.819 &   0.214 &   2.371 &    52.0 &    28.6 &    77.1 &    10.37 &     0.61 &     0.51 &     1.13 &   -52.2 &    15.2 &   0.96 &  \myeol 
 3996137902634436480 &      641 &      578 &      717 &   0.820 &   0.547 &   1.122 &    38.5 &    34.7 &    42.1 &    39.71 &     1.07 &    10.51 &     2.73 &   -38.4 &     2.3 &   0.95 &  \myeol 
 3260079227925564160 &      909 &      890 &      928 &   0.824 &   0.798 &   0.849 &    33.4 &    32.7 &    34.1 &    32.19 &     0.06 &     6.00 &     0.12 &   -33.4 &     0.4 &   0.47 &  \myeol 
 5700273723303646464 &    -1636 &    -1846 &    -1464 &   0.836 &   0.227 &   1.768 &    38.1 &    36.6 &    39.5 &    15.70 &     1.08 &     0.26 &     1.98 &    38.0 &     0.9 &   0.95 &  $\dagger$ \myeol 
 5551538941421122304 &    -3815 &    -4023 &    -3626 &   0.866 &   0.772 &   0.955 &    30.4 &    28.8 &    31.9 &     8.47 &     0.01 &     1.03 &     0.04 &    30.0 &     1.0 &   0.65 &  \myeol 
 2924378502398307840 &    -1841 &    -1880 &    -1803 &   0.880 &   0.803 &   0.957 &    87.1 &    85.5 &    88.8 &     6.10 &     0.03 &     0.74 &     0.06 &    87.0 &     1.0 &   0.75 &  \myeol 
 6724929671747826816 &     1045 &      986 &     1111 &   0.884 &   0.512 &   1.324 &    54.8 &    53.0 &    56.7 &    17.07 &     0.49 &     3.13 &     0.99 &   -54.8 &     1.1 &   0.72 &  \myeol 
 3972130276695660288 &     -511 &     -530 &     -493 &   0.888 &   0.855 &   0.921 &    31.9 &    30.7 &    33.0 &    59.97 &     0.05 &    22.01 &     0.14 &    31.8 &     0.7 &   0.58 &  \myeol 
 5163343815632946432 &    -4965 &    -5499 &    -4523 &   0.896 &   0.484 &   1.361 &    37.1 &    33.7 &    40.7 &     5.39 &     0.04 &     1.45 &     0.09 &    35.4 &     2.3 &   0.76 &  \myeol 
 2926732831673735168 &    -1680 &    -1698 &    -1664 &   0.917 &   0.820 &   1.014 &    66.5 &    66.1 &    66.9 &     8.75 &     0.04 &     0.99 &     0.08 &    66.5 &     0.3 &   1.15 &  \myeol 
 2929487348818749824 &    -5364 &    -5645 &    -5108 &   0.926 &   0.239 &   2.004 &    70.0 &    67.9 &    72.3 &     2.61 &     0.06 &     0.42 &     0.10 &    69.9 &     1.4 &   1.34 &  \myeol 
  939821616976287104 &      -90 &      -91 &      -90 &   0.989 &   0.974 &   1.002 &   568.4 &   567.1 &   569.6 &    19.05 &     0.07 &    45.71 &     0.11 &   568.2 &     0.8 &     NA & $\dagger$ \myeol 
 3458393840965496960 &     -866 &    -1419 &     -600 &   0.996 &   0.396 &   2.104 &    86.4 &    52.8 &   120.9 &    13.20 &     1.05 &     2.79 &     2.06 &    86.6 &    19.9 &   1.17 &  \myeol 

\bottomrule
\end{tabular}
}
\end{table*}

\begin{table*}
\centering
\tiny{
\caption{
Additional data from \gdr{2} for the close encounters listed in Table~\ref{tab:periStats}. They are all taken directly from the catalogue,
except for the astrometric ``unit weight error'' $u$ which is calculated as {\tt sqrt[astrometric\_chi2\_al/(astrometric\_n\_good\_obs\_al-5)]}. 
The online table at CDS includes all 3379 stars in the filtered sample.
\label{tab:extras}}
\tabcolsep=0.14cm
\begin{tabular}{*{10}{r}}
\toprule
\gdr{2} source ID & G       & BP-RP & $u$ & No.\       & astrometric & astrometric & No.\   & $l$ & $b$ \\
                          & mag  & mag     &       & visibility  & excess      & excess           &  RVS   & deg & deg\\
                          &         &            &        & periods   & noise          & noise sig       & transits &  & \\
\midrule
 4270814637616488064 &  9.06 &  1.70 &   1.22 &  10 &     0.00 &     0.00 &  2 &  27 &   6 \myeol 
  955098506408767360 & 12.41 &  0.76 &  16.19 &  10 &     2.49 &  1661.28 &  5 & 176 &  10 \myeol 
 5571232118090082816 & 11.79 &  1.50 &   1.25 &  15 &     0.00 &     0.00 & 10 & 249 & -25 \myeol 
 2946037094755244800 & 12.34 &  1.49 &  27.41 &  10 &     3.12 &  4225.87 & 22 & 228 &  -7 \myeol 
 4071528700531704704 & 12.44 &  0.78 &  23.52 &  10 &     3.92 &  3138.25 &  3 &   7 & -11 \myeol 
  510911618569239040 &  8.88 &  0.77 &   1.57 &  17 &     0.00 &     0.00 &  6 & 126 &   0 \myeol 
  154460050601558656 & 15.37 &    NA &   6.47 &  10 &     1.85 &   359.52 &  3 & 174 & -11 \myeol 
 6608946489396474752 & 12.28 &  1.44 &   0.89 &  10 &     0.00 &     0.00 &  6 &  23 & -61 \myeol 
 3376241909848155520 & 12.52 &  0.78 &  28.53 &  11 &     5.46 &  8824.25 &  8 & 190 &   5 \myeol 
 1791617849154434688 & 11.00 &  1.09 &   1.49 &  11 &     0.00 &     0.00 &  9 &  70 & -18 \myeol 
 4265426029901799552 & 12.20 &  2.07 &  18.88 &  10 &     3.44 &  2533.02 &  3 &  32 &   0 \myeol 
 5261593808165974784 & 12.69 &  2.02 &   1.39 &  18 &     0.00 &     0.00 & 14 & 285 & -27 \myeol 
 5896469620419457536 & 13.55 &  1.98 &   1.06 &  11 &     0.03 &     0.23 &  5 & 313 &   6 \myeol 
 4252068750338781824 & 12.10 &  0.92 &  20.42 &   9 &     2.53 &  1453.10 &  5 &  26 &  -3 \myeol 
 1949388868571283200 & 13.12 &    NA &   1.34 &  11 &     0.13 &     5.81 &  2 &  86 & -13 \myeol 
 1802650932953918976 & 12.69 &  1.02 &   0.99 &  14 &     0.00 &     0.00 & 11 &  52 & -11 \myeol 
 3105694081553243008 & 12.31 &  1.24 &  25.05 &   9 &     4.78 &  6025.61 &  6 & 215 &  -2 \myeol 
 5231593594752514304 & 12.03 &  1.87 &   1.22 &  16 &     0.00 &     0.00 &  2 & 293 &  -9 \myeol 
 4472507190884080000 & 12.90 &  0.87 &  13.00 &   9 &     2.17 &  1206.88 &  5 &  29 &  15 \myeol 
 3996137902634436480 & 11.74 &  0.87 &  32.65 &   9 &     5.07 &  7026.13 &  4 & 212 &  64 \myeol 
 3260079227925564160 & 11.73 &  2.13 &   1.98 &  13 &     0.12 &     8.12 &  7 & 188 & -33 \myeol 
 5700273723303646464 & 11.96 &  0.84 &  32.64 &  15 &     5.26 &  8991.52 &  8 & 242 &   7 \myeol 
 5551538941421122304 & 13.10 &  1.71 &   1.23 &  16 &     0.06 &     1.31 & 11 & 257 & -21 \myeol 
 2924378502398307840 & 12.62 &  1.25 &   1.10 &  14 &     0.00 &     0.00 & 14 & 232 & -16 \myeol 
 6724929671747826816 & 11.97 &  1.31 &  13.55 &   9 &     1.87 &  1105.66 &  2 & 352 & -11 \myeol 
 3972130276695660288 &  9.88 &  2.18 &   1.58 &   8 &     0.00 &     0.00 &  3 & 227 &  65 \myeol 
 5163343815632946432 & 12.92 &  1.25 &   1.21 &  10 &     0.00 &     0.00 &  5 & 194 & -50 \myeol 
 2926732831673735168 &  9.56 &  0.72 &   1.41 &  14 &     0.00 &     0.00 &  9 & 230 & -12 \myeol 
 2929487348818749824 & 11.21 &  0.78 &   1.54 &  12 &     0.00 &     0.00 &  4 & 233 &  -5 \myeol 
  939821616976287104 &  9.91 &  1.43 &   1.50 &   8 &     0.00 &     0.00 &  2 & 181 &  18 \myeol 
 3458393840965496960 & 12.07 &  0.63 &  11.71 &   9 &     3.07 &  2046.73 &  2 & 172 &   9 \myeol 

\bottomrule
\end{tabular}
}
\end{table*}

\begin{figure}
\begin{center}
\includegraphics[width=0.5\textwidth, angle=0]{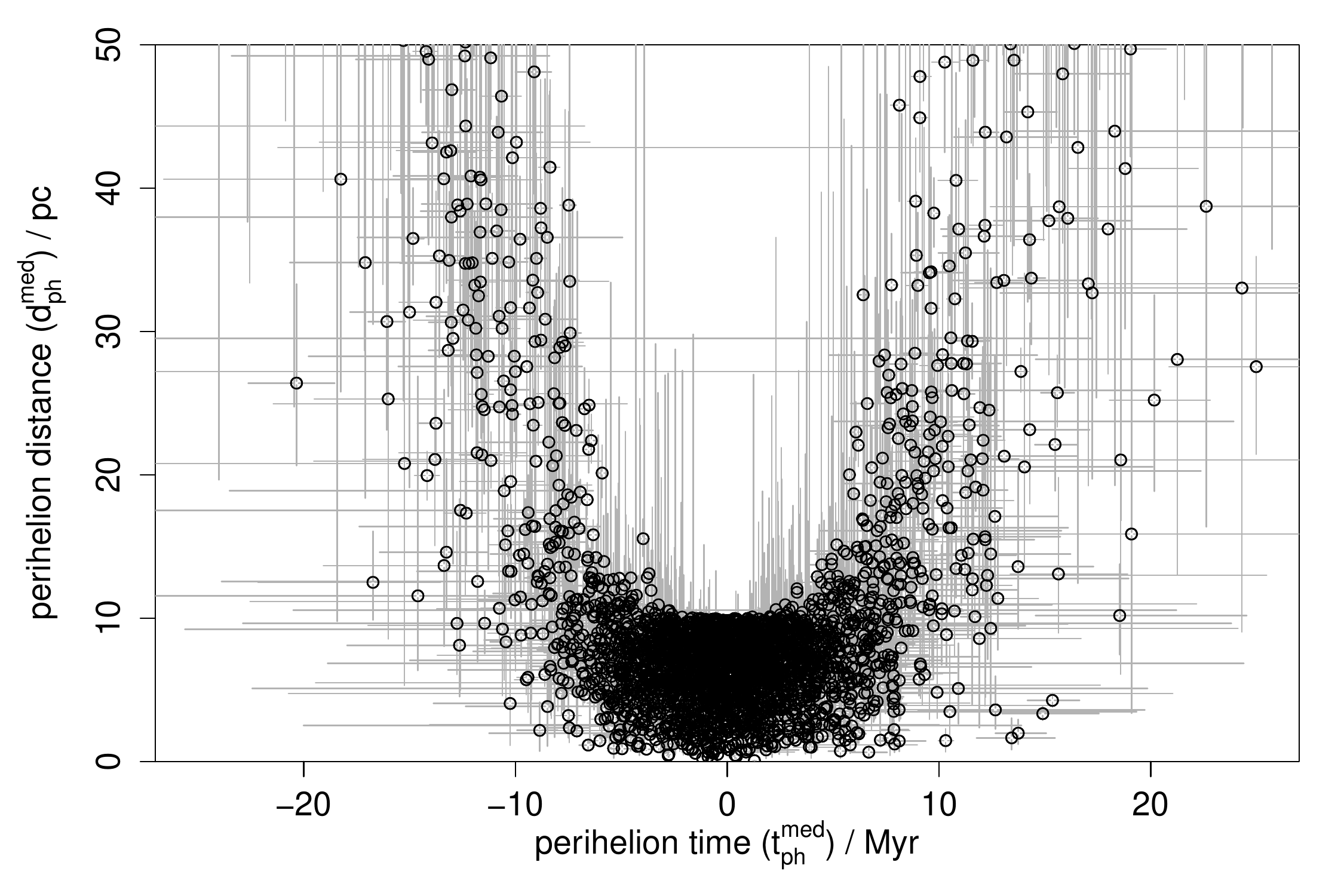}
\caption{Perihelion times and distances computed by orbit integration for the filtered sample (278 points lie outside the plotting range). Open circles show the median of the perihelion time and distance distributions. The error bars show the limits of the 5\% and 95\% percentiles.
\label{fig:dph_vs_tph_0to50pc_witherrors}}
\end{center}
\end{figure}

\begin{figure}
\begin{center}
\includegraphics[width=0.5\textwidth, angle=0]{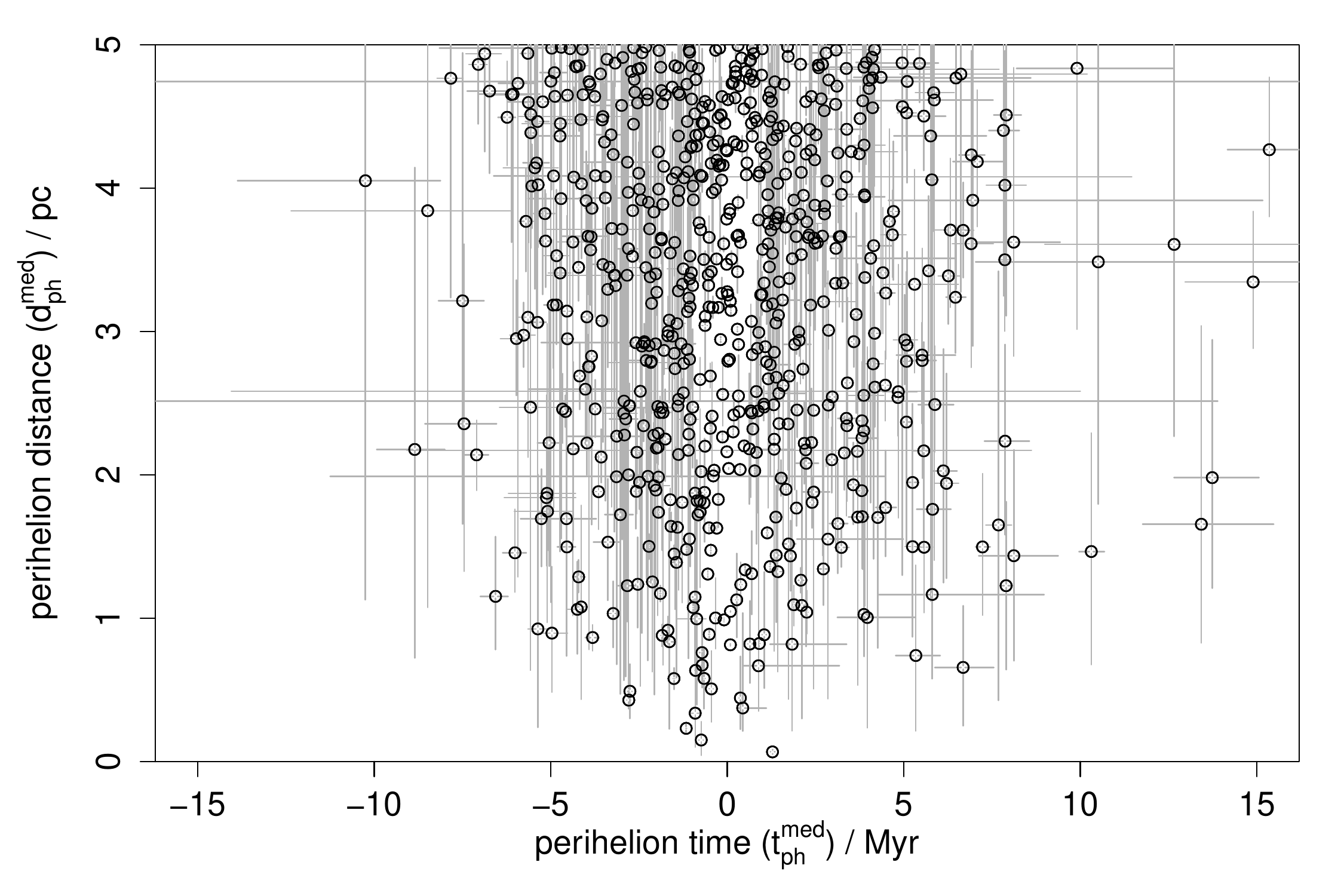}
\caption{As Figure \ref{fig:dph_vs_tph_0to50pc_witherrors}, but just showing the 694 stars with $\dphmed<5$\,pc.
The time axis is scaled to show all encounters in this perihelion distance range.
\label{fig:dph_vs_tph_0to5pc_witherrors}}
\end{center}
\end{figure}

Figure~\ref{fig:dph_vs_tph_0to50pc_witherrors} plots the perihelion times and distances. 
(This and other plots do not remove bogus cases.)
The stars were selected by the LMA to come within 10\,pc of the Sun, but some of the orbit integrations result in much larger median perihelion distances.
Similarly, our list will be missing those stars which would have 
had $\dphmed<10$\,pc had they been subject to the orbital integration, but were never selected because they had
$\dph>10$\,pc from the LMA. 
This latter omission will in principle lead to an underestimate of the derived encounter rate.
However, as shown in both papers 1 and 2, this mostly affects stars that will encounter further in the past/future and/or nearer to the edge of the 10\,pc distance limit.
As stars at such large distances can hardly be considered encountering,
we will only be interested in encounters within 5\,pc from now on. Moreover, when we later compute the (completeness-corrected) encounter
rate, we will limit the sample to a narrower time window. 

The encounters with $\dphmed < 5$\,pc are shown in Figure~\ref{fig:dph_vs_tph_0to5pc_witherrors}.
We see a strong drop in the density of encounters with increasing $|\tph|$. This is primarily a consequence of the magnitude limit in the sample (95\% brighter than \gmag\,=\,13.4\,mag): 
Encounters that would occur further in the past/future generally correspond to stars that are currently more distant, and so more likely to be below the limiting magnitude. The effective time limit of this study is 5--10\,Myr. Comparing this with Figure 3 of \citetalias{2018A&A...609A...8B}, we see that while \gdr{2} has found many more encounters than our TGAS study (by a factor of about seven within 5\,pc), \gdr{2} does not allow us to probe much further into the past/future, because of the similar magnitude limits on the samples.

We also see in Figure~\ref{fig:dph_vs_tph_0to5pc_witherrors} a slight reduction in the density of encounters very close to the present time. (The effect is not quite as strong as first appears, as it is partly an illusion produced by the shorter error bars.)
This was also seen with in \citetalias{2018A&A...609A...8B}, where we argued this was due to two things: missing bright stars in the \gaia\ catalogue; and the ever smaller volume available for encounters to occur at arbitrarily near times. Both of these apply for \gdr{2}, although we argue later that this is also a consequence of the 
limited \teff\ range for stars with radial velocities in \gdr{2}.

\begin{figure}
\begin{center}
\includegraphics[width=0.5\textwidth, angle=0]{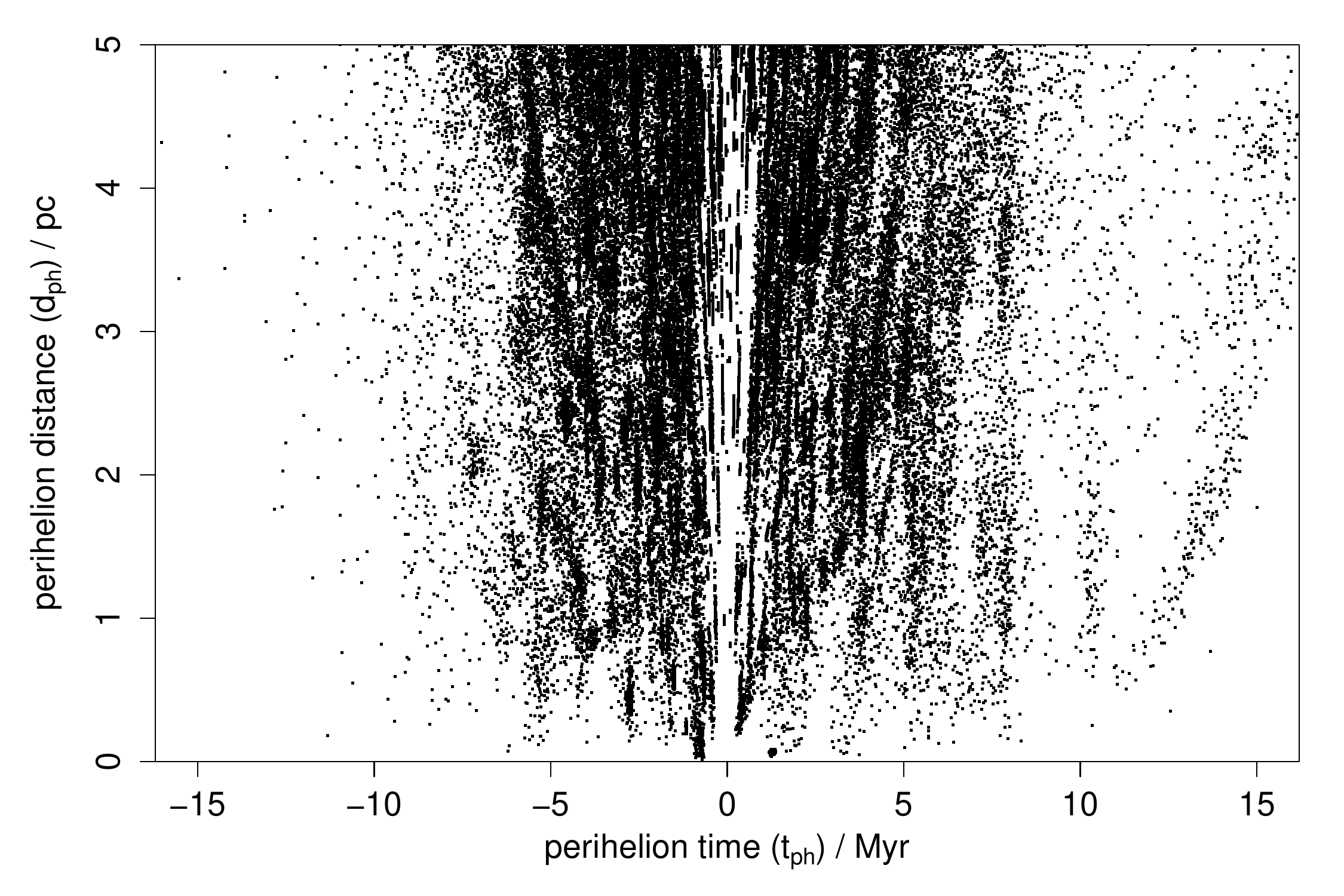}
\caption{As Figure \ref{fig:dph_vs_tph_0to5pc_witherrors}, but now showing the individual surrogates from the orbital integrations (but just plotting 100 per star rather than all 2000 per star). Surrogates from stars with median perihelion parameters outside the plotting range (and so not shown in Figure \ref{fig:dph_vs_tph_0to5pc_witherrors}) {\em are} shown here.
\label{fig:dph_vs_tph_0to5pc_samples}}
\end{center}
\end{figure}

The uncertainties in $\tph$ and $\dph$ are correlated. This can be seen in Figure~\ref{fig:dph_vs_tph_0to5pc_samples}, where we plot the individual surrogates instead of the summary median and axis-parallel error bars. Each star is generally represented as an ellipsoidal shape pointing roughly towards $\tph=0$, $\dph=0$.  

\begin{figure}
\begin{center}
\includegraphics[width=0.5\textwidth, angle=0]{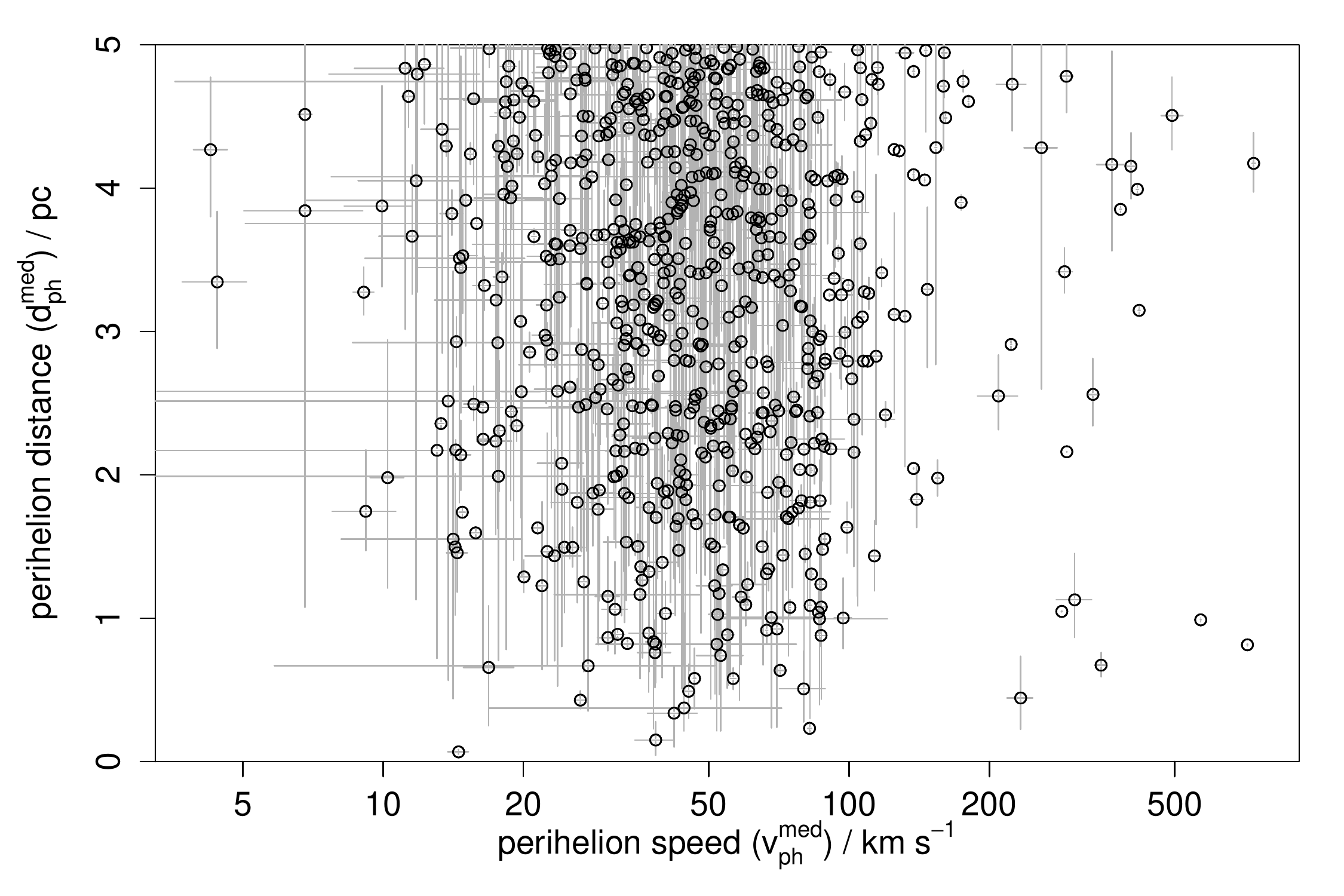}
\caption{Median perihelion velocities from the orbit integrations for those encounters shown in Figure \ref{fig:dph_vs_tph_0to5pc_witherrors} The velocity axis is a logarithmic scale.
\label{fig:dph_vs_vph_0to5pc_witherrors}}
\end{center}
\end{figure}

The perihelion speeds for encounters with $\dphmed<5$\,pc are shown in Figure \ref{fig:dph_vs_vph_0to5pc_witherrors}.  90\% have $\vphmed<100$\,\kms.
The very fast encounters are almost entirely due to stars with large radial velocities. 

The masses of the encountering stars are listed in Table~\ref{tab:periStats}. The 5th, 50th, and 95th quantiles of the mass distribution over the entire filtered sample are
0.53, 0.91, and 1.64\,\Msol\ respectively. The lack of massive stars in the sample is a consequence of the \teff\ filtering on radial velocities published in \gdr{2}.

\begin{figure}
\begin{center}
\includegraphics[width=0.5\textwidth, angle=0]{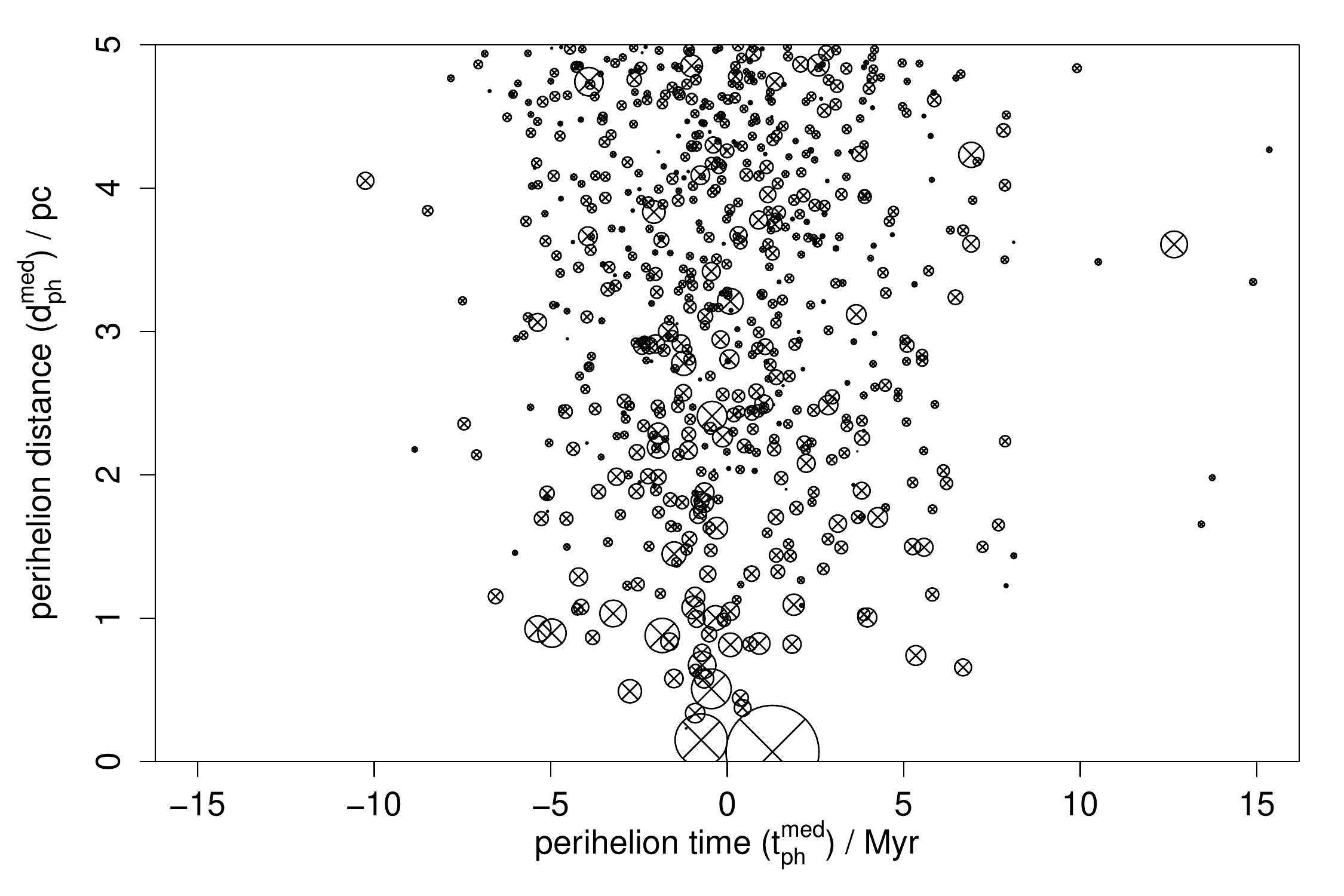}
\caption{As Figure \ref{fig:dph_vs_tph_0to5pc_witherrors}, but 
now plotting each star as a circle, the area of which is proportional to 
$\mass/(\vphmed \dphmed)$.
\label{fig:dph_vs_tph_areaisiimpulse1}}
\end{center}
\end{figure}

\begin{figure}
\begin{center}
\includegraphics[width=0.5\textwidth, angle=0]{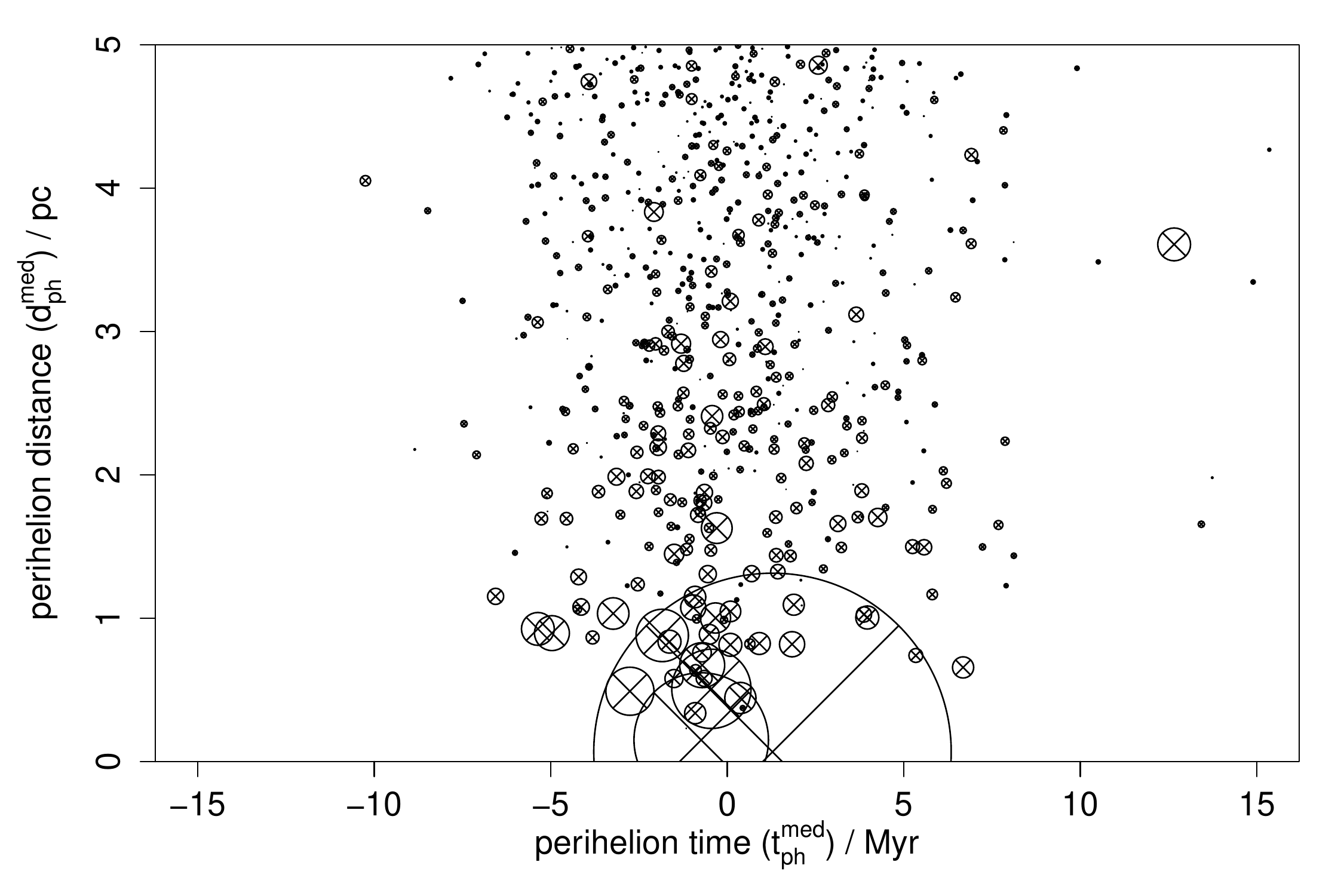}
\caption{As Figure \ref{fig:dph_vs_tph_0to5pc_witherrors}, but 
now plotting each star as a circle, the area of which is proportional to 
$\mass/(\vphmed (\dphmed)^2)$.
\label{fig:dph_vs_tph_areaisiimpulse2}}
\end{center}
\end{figure}

Close encounters are interesting not least because they perturb the Oort cloud. The degree of perturbation, or rather the impulse which comets can receive, depends not only on the distance of the encountering star, but also on its speed and the mass.
According to the simple impulse approximation 
\citep{10.2307/20022899, 1950BAN....11...91O, 1976BAICz..27...92R, 1994CeMDA..58..139D}
the impulse transfer is given by
$M \vph^{-1} \dph^{-\alpha}$ where $\alpha$ is 1 for very close encounters (on the order of the comet--Sun separation), and 2 otherwise.
The impulse of the encountering stars is visualized in Figures~\ref{fig:dph_vs_tph_areaisiimpulse1} and Figures~\ref{fig:dph_vs_tph_areaisiimpulse2} for $\alpha$ equal to 1 and 2 respectively.
Regardless of which impulse approximation we use, it is the closest encounters which have the greatest impact.

\subsection{Colour--magnitude diagram}\label{sec:cmd}

\begin{figure}
\begin{center}
\includegraphics[width=0.5\textwidth, angle=0]{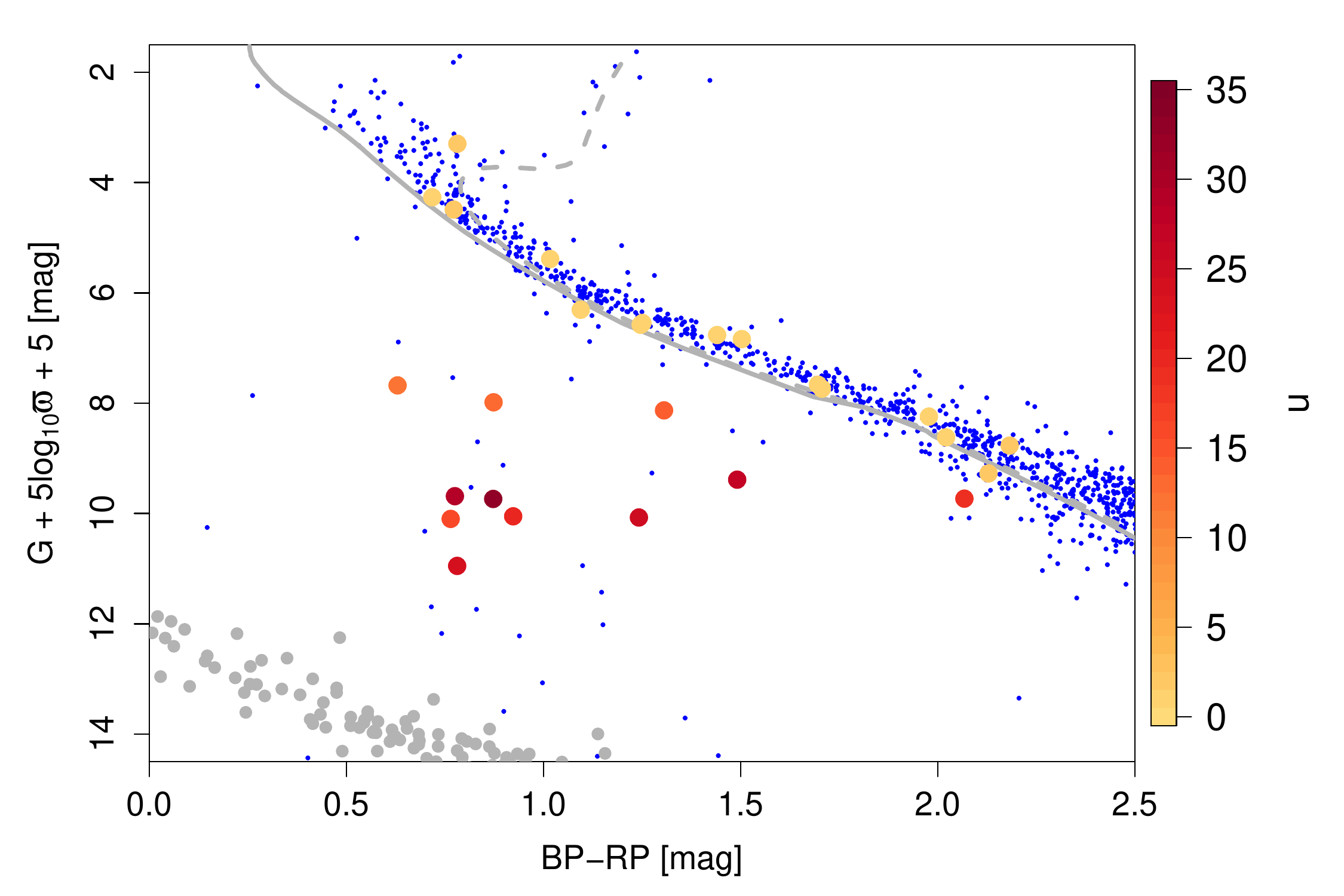}
\includegraphics[width=0.5\textwidth, angle=0]{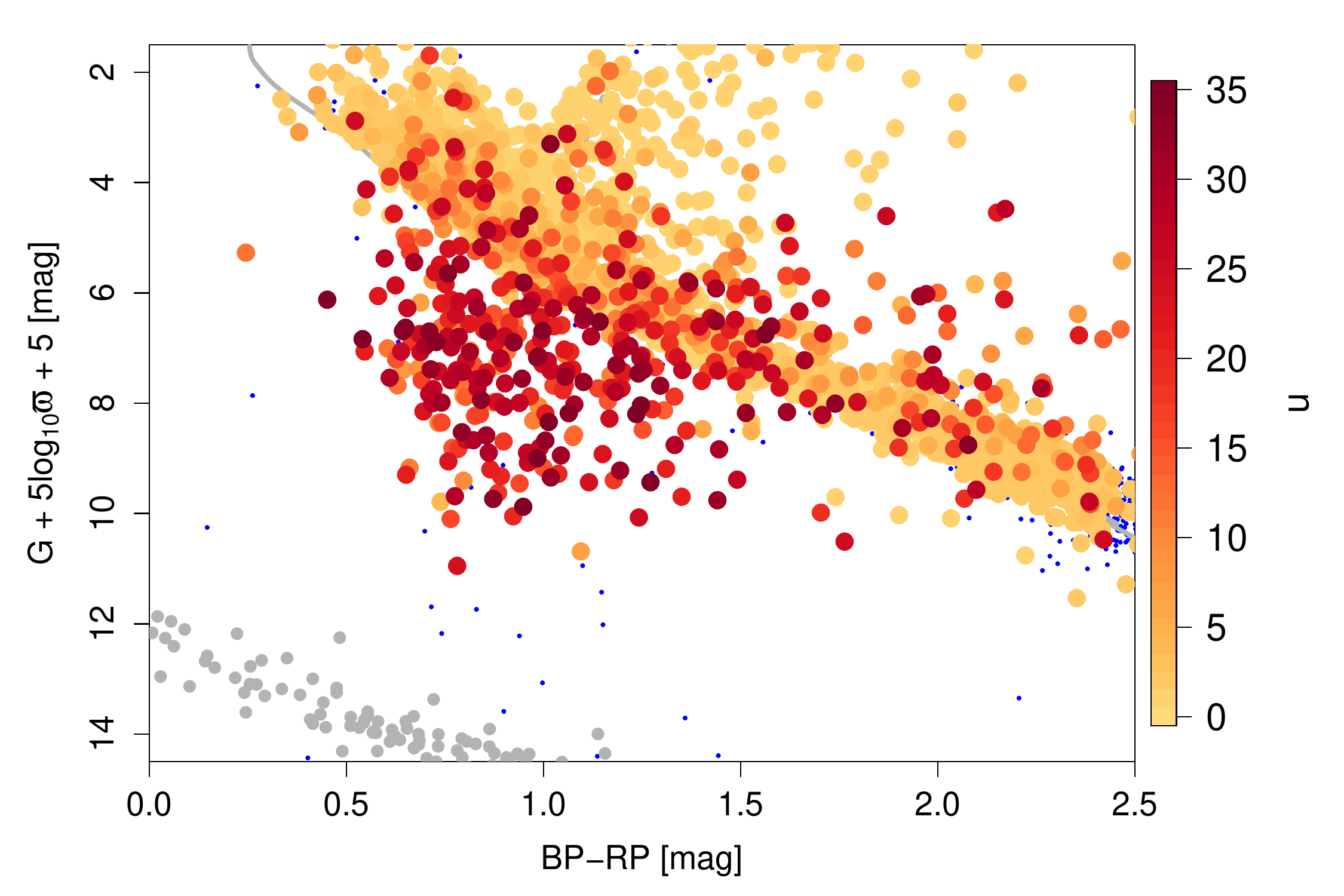}
\caption{Colour--magnitude diagram for stellar encounters coloured according to the 
unit weight error $u$ in Table~\ref{tab:extras}. The vertical axis equals $M_G$ assuming extinction is zero. For orientation, the grey lines are unreddened solar metallicity PARSEC isochrones for 1\,Gyr (solid) and 10\,Gyr (dashed) from \cite{2017ApJ...835...77M}, 
the grey points are white dwarfs within 20\,pc of the Sun identified by
\cite{2018arXiv180512590H} (many lie outside the range of the plot), and the
small blue dots show all sources in \gdr{2} with $\parallax>50$\,mas (plotted on the bottom layer, so most are obscured by coloured and grey points, especially in the lower panel).
The upper panel shows all reliable (non-bogus) encounters with $\dphmed<1$\,pc (from Table~\ref{tab:periStats}). For comparison, the lower panel shows all encounters in the filtered sample (some lie outside the plotting range). Encounters have been plotted with larger values of $u$ on higher layers so as to better locate larger values of $u$. The colour bar spans the full range of values in both panels.
\label{fig:cmd}}
\end{center}
\end{figure}

Figure~\ref{fig:cmd} shows the colour--absolute magnitude diagram (CMD) for the close encounters on the assumption of zero interstellar extinction.
The upper panel shows the 26 non-bogus encounters with $\dphmed<1$\,pc from Table~\ref{tab:periStats}. Eleven of these have unit weight error $u>10$ (the others have $u<2$) and lie some way below and left of the main sequence (MS), yet well above the white dwarf (WD) sequence.
The dimensionless quantity $u$ measures how well the five-parameter solution describes the astrometric data (see section~\ref{sec:filtering}). 
If we assumed that all the encounters were single main sequence stars, then one
interpretation of this diagram is that the parallaxes of these 11 sources are wrong; specifically that they are all overestimated by one or two orders of magnitude (i.e.\ the sources should be much further away). This seems unlikely, although it is the case that a search for encounters will preferentially include sources with spuriously large parallaxes rather than sources with spuriously small parallaxes (because the former are more likely to have apparent close encounters). 
As we see a smaller proportion of sources between the MS and WDs among nearby stars (blue points in Figure~\ref{fig:cmd}), it at first seems unlikely that we would get such a large proportion of such sources among the closest encounters (upper panel).
The magnitude of this selection effect is impossible to estimate without knowing which parallaxes are actually spurious.
On the other hand, if we look instead at {\em all} encounters (lower panel), then we find that just 14\% of all encounters 
within the plotted range lie below the MS, whereas this figure is 24\% for the nearby stars. 
In other words, there is in fact a decreased tendency for encounters to be below the MS compared to nearby stars.
This comparison is not ideal, however, because 
some of these nearby stars are WDs -- essentially absent from the encountering sample due to the magnitude limit -- and the encounters are drawn from a much larger volume

We noted in section~\ref{sec:filtering} that a large value of $u$ does not mean the parallax is wrong. Indeed, a relatively poor fit of the five-parameter solution is in principle expressed by larger values of the astrometric uncertainties (and these are quite correlated with $u$). Here, the 11 sources in the upper panel of  Figure~\ref{fig:cmd} have $\sigma(\parallax)$ between 0.5\, and 1.1\,mas, whereas the other 15 have $\sigma(\parallax)<0.06$\,mas.
We have already accommodated the uncertainties in the identification of the close encounters.  Furthermore, although the parallax uncertainties are larger, they produce uncertainties in the absolute magnitude of no more than 0.08\,mag, which is smaller than the size of the points plotted in Figure~\ref{fig:cmd}. Contamination of the astrometric solution by nearby (sub-arcsecond) sources is possible, 
and although most of the sources are near to the Galactic plane (as are unquestioned sources), they are all bright, so less likely to be affected by crowding.
Visual inspection of non-Gaia images of these targets shows no obvious contaminants.
They do generally have fewer visibility periods in the solutions, however: 9, 10, or 11, as opposed to 8--18 for the sources lying near the main sequence. 

Given their location in the CMD, it is possible that some or all of these 11 sources are subdwarfs or MS-WD binaries.
The latter, of which many have been discovered (e.g.\ \citealt{2016MNRAS.458.3808R}), can in principle lie in much of the region between the MS and WD sequence. Such binarity could potentially explain elevated values of $u$, although only if the periods were short enough and the amplitudes large enough to be detectable over \gdr{2}'s 22-month baseline. Even then the parallax should be reasonably accurate, because the different astrometric displacements at different times should not create a large bias (Lennart Lindegren, private communication). Furthermore, four of the 11 sources have much larger radial velocity uncertainties than the other encounters in Table~\ref{tab:periStats} ($>14$\,\kms as opposed to $<6$\,\kms). As this uncertainty is computed from the standard deviation of  measurements at different epochs (equation 1 of \citealt{2018arXiv180409372K}),
this could be an indication of binarity.

In conclusion, the correlation of the CMD location with $u$ is suspicious, and would be consistent with a selection effect that preferentially includes spuriously large parallaxes in a search for stellar encounters. On the other hand, we do not have specific evidence to claim that these stars have spurious solutions or unaccounted for errors, and it is quite possible that some or all are MS-WD binaries.  Spectroscopy may help identify the nature of these sources, and astrometry over a longer timebase (e.g.\ from \gdr{3}) should help to determine if the astrometric solutions are good.

\subsection{Individual encounters found to be coming within 1\,pc}

We find 31 stars with $\dphmed<1$\,pc. All have been visually checked against 2MASS, DSS, PS1, and AllWISE images in Aladin and all are confirmed as real sources. 
Some have moderately high values of $u$, 
{\tt astrometric\_excess\_noise} and/or {\tt astrometric\_excess\_noise\_sig}, as can be seen in Table~\ref{tab:extras}.
Only one of the encounters has been discovered by the previous studies listed in the introduction.
21 do not have Hipparcos or Tycho IDs, so could not have been discovered in papers 1 or 2.
Of the remaining nine stars, seven have obvious Tycho matches but were not found in paper 2 for reasons discussed below. The last two stars are problematic and are also discussed below.

The closest approaching star, and the only one to have been found already, is
\object{\gdr{2} 4270814637616488064}, better known as
\object{Gl 710} (\object{Tyc 5102-100-1}, \object{Hip 89825}). This is
a K7 dwarf known from many previous studies to be a very close encounter. 
In \citetalias{2015A&A...575A..35B} we found 
a median encounter distance of 0.267\,pc (90\% CI 0.101--0.444\,pc). 
A much smaller -- and more precise -- proper motion measured by TGAS gave a closer encounter, 
with a median distance of 0.076\,pc (90\% CI 0.048--0.104\,pc) (\citetalias{2018A&A...609A...8B} and \citealt{2016A&A...595L..10B}).
\gdr{2} has slightly decreased this distance -- 0.0676\,pc (13\,900\,AU) -- and has narrowed the uncertainty -- 90\% CI  0.0519--0.0842\,pc (10\,700--17\,400\,AU) --
although this is statistically consistent with the TGAS result. 
This is well within the Oort cloud.
Due to the slightly more negative radial velocity published in \gdr{2} than used in the earlier studies, the encounter time is slightly earlier (but no more precisely determined). 
We note, however, that the radial velocity in \gdr{2} comes from just two \gaia\ focal plane transits, the minimum for a value to be reported in the catalogue. Despite its low mass, \object{Gl 710}  imparts the biggest impulse according to either impact approximation (Figures~\ref{fig:dph_vs_tph_areaisiimpulse1} and~\ref{fig:dph_vs_tph_areaisiimpulse2}), in part because of its low velocity.
Using the same data, \cite{2018arXiv180502644D} get a slightly closer mean (and median) distance of 0.052\,pc (with a symmetric standard deviation of 0.010\,pc).
Their estimate is close to what one gets for a constant gravitational potential, namely
0.0555\,pc (computed by propagating the set of surrogates using the LMA and taking the mean).
This suggests they used a very different potential from ours.
Note that ignoring the parallax zeropoint 
and/or ignoring the astrometric correlations 
changes the estimate by less than 0.0001\,pc, so these are not the cause of the difference.

The second closest approach in Table~\ref{tab:periStats}, \object{\gdr{2} 955098506408767360}, is also the one with the second largest impulse. It is one of the most massive of the closest encounters, with a perihelion in the relatively recent past.
Four fainter 2MASS sources lie within 8$"$, but there is no good reason to think these have interfered with the \gaia\ solution.
 The most recent reliable encounter in the table is  
\object{\gdr{2} 3376241909848155520}, which has a 50\% chance of having passed within 0.5\,pc, around 450\,kyr ago.

Among the 21 sources without Tycho matches, most have matches to 2MASS (and a couple of others to other surveys). A few of these are in very crowded regions (e.g\ at low Galactic latitude) and have close companions, but in all but one case these companions are much fainter, and there is no specific evidence to suggest a problem with the \gaia\ measurements. The one exception is \object{\gdr{2} 5700273723303646464}, which has a companion 4$"$ away that is just 1.6\,mag fainter in the J-band (the best match to this in \gdr{2} is 0.7\,mag fainter in the G band, and has a completely different parallax and proper motion).
This is a borderline case. Its measurements are not suspicious, but some of the quality metrics make it questionable. We decide to flag this as bogus.

There are seven encounters with obvious Tycho matches that were not found as encounters in paper 2, either because they lacked a radial velocity, or because their astrometry has changed substantially from TGAS to \gdr{2}. These are as follows (with notes on the problematic cases).

\begin{itemize}
\item \object{\gdr{2} 510911618569239040} = \object{TYC 4034-1077-1} .
\item \object{\gdr{2} 154460050601558656} = \object{TYC 1839-310-1}. This is rather faint, $\gmag=15.37$\,mag, and has a rather high radial velocity based on just 3 transits. It also has no BP-RP colour and a large {\tt astrometric\_excess\_noise\_sig} of 360, so we consider this encounter to be bogus.
\item \object{\gdr{2} 1791617849154434688} = \object{TYC 1662-1962-1}.
\item \object{\gdr{2} 1949388868571283200} = \object{TYC 2730-1701-1}. The large radial velocity is based on just two transits, so we consider this to be bogus.
\item \object{\gdr{2} 3972130276695660288} = \object{TYC 1439-2125-1} = \object{GJ 3649}.
\item \object{\gdr{2} 2926732831673735168} = \object{TYC 5960-2077-1}.
\item \object{\gdr{2} 2929487348818749824} = \object{TYC 5972-2542-1}.
\end{itemize}

The final two sources have ambiguous matches. Simbad lists four objects within 5$"$ of \object{\gdr{2} 5231593594752514304} ($\gmag=12.03$\,mag), 
although in reality these four are probably just two or three unique sources. 
\gdr{2} itself identifies one of these as
\object{\gdr{2} 5231593599052768896}, which
with $\gmag=9.12$\,mag is almost certainly \object{Hip 53534} with $V=9.21$\,mag.
Whichever other star our close encounter is, its RVS spectrum is surely contaminated by this much brighter star just a few arcseconds away, suggesting that the whoppingly large radial velocity of $-715$\,\kms\ measured by RVS is in fact spurious. We therefore consider this encounter to be bogus.

Finally, \object{\gdr{2} 939821616976287104} would appear to match with \object{TYC 2450-1618-1}. However the 2MASS image clearly shows this to be a double star, and this surely has contaminated the RVS spectrum. In that case we do not trust the high radial velocity reported (568\,\kms) and conservatively consider this to be a bogus encounter too.

\subsection{Individual encounters coming within 1\,pc in papers 1 and 2, but not found in the present study}

In paper~2 we found two stars with $\dphmed<1$\,pc, one of which was \object{Gl 710}, discussed above. The other was 
\object{Tyc 4744-1394-1} which is in \gdr{2} but without a radial velocity. The astrometry is consistent with that from TGAS, confirming the reality of that encounter also.

In paper~1 we found 14 stars coming within 1\,pc. One was \object{Gl 710}. The other 13 are not found in the present study for various reasons. We briefly discuss each of these cases. The match in \gdr{2} was found by searching for sources within about 30$"$ using Simbad, and identifying the one with the closest magnitude. In all cases but one the match was unambiguous, and in none of these is there anything to suggest the \gdr{2} data are problematic.

\object{Hip 85605}.
This was the closest encounter found in \citetalias{2015A&A...575A..35B} 
at $\dphmed=0.103$\,pc (90\% CI 0.041--0.205\,pc). It was deemed dubious due to the inconsistency of its magnitude and Hipparcos parallax with its apparent spectral type. This suspicion is confirmed with \gdr{2}, which has a much smaller parallax of $1.822\pm0.027$\,mas
compared to $146.84\pm29.81$ in Hipparcos-2 (the proper motions and radial velocities are consistent within 1$\sigma$).
This now places the closest approach at 420\,pc. It was conjectured in paper~1 that the cause for the incorrect 
parallax is a binary companion at a very problematic separation for the Hipparcos detectors.

\object{Hip 63721}. Its best match, \object{\gdr{2}  1459521872495435904}, has a parallax of $5.99 \pm 0.04$\,mas, compared to 
$216.62 \pm 56.53$\,mas in Hipparcos. This Hipparcos solution was doubted in paper~1 on the grounds of inconsistency with spectral type and it being a double star.

\object{Hip 91012}. The parallax and proper motion of its best match, \object{\gdr{2}  4258121978471892864}, are consistent with Hipparcos, but the \gdr{2} radial velocity is now much smaller, at  $-16.27 \pm 0.61$\,\kms, compared to the RAVE value of $-364.1 \pm 22.4$ adopted in paper~1.

\object{Hip 23311}. This matches to \object{\gdr{2}  3211461469444773376}. It is a high proper motion ($> 1000$\,\maspyr) nearby ($113$\,mas) star -- \gdr{2} and Hipparcos agree -- but the \gdr{2} radial velocity is much smaller: $21.38 \pm 0.15$\,\kms\ compared to the implausibly large value of $-813.7 \pm 48.6$ from RAVE that led it to being flagged as dubious in paper~1.

\object{Hip 85661}. The parallax and radial velocity used in paper~1 agree with the best match star, \object{\gdr{2} 4362793767434602752}, but the \gdr{2} proper motion is much larger ($8.5 \pm 0.2$\,\maspyr\ versus $0.5 \pm 0.6$\,\maspyr), meaning it no longer comes near to the Sun.

\object{Hip 55606}. The proper motion of its best match, \object{\gdr{2}  3590767623539652864}, is consistent with Hipparcos (parallax less so), but the \gdr{2} radial velocity is now much smaller, at $-18.12 \pm 0.70$\,\kms, compared to the implausibly large value of $-921.0 \pm 91.9$ from RAVE that led it to being flagged as dubious in paper~1.

\object{Hip 75159}. Its counterpart, \object{\gdr{2}  6251546901899531776}, has no radial velocity in \gdr{2}. 
The \gdr{2} astrometry agrees within 2$\sigma$ of the Hipparcos astrometry (which was rather imprecise, with a proper motion consistent with zero).
Using the (large) RAVE radial velocity of 368\,\kms\ from paper 1 with the \gdr{2} astrometry, the LMA gives a closest approach of 2.1\,pc 660\,kyr in the past.

\object{Hip 103738} (\object{gamma Microscopii}). This G6 giant 
was the potentially most massive encounter coming within 1\,pc in paper~1,
found to have $\dphmed=0.83$\,pc (90\% CI 0.35--1.34\,pc) based on the Hipparcos-2 proper motion of $1.78\pm0.35$\,\maspyr.  (A second entry in the catalogue based on the larger Tycho-2 proper motion put it at 3.73\,pc; 90\% CI 2.28--5.22.)
This star is \object{\gdr{2} 6781898461559620480} with $\gmag=4.37$\,mag. The proper motion in \gdr{2} is ten times (17$\sigma$) larger, $17.7\pm0.9$\,\maspyr, with no suggestion from the quality metrics that this is a poor solution. (Both the parallaxes and radial velocities agree with 1$\sigma$.)  Either there is a significant, unaccounted for acceleration, or the proper motion in Hipparcos-2 or \gdr{2} (or both) is wrong.
Using the \gdr{2} data we find the encounter to be at 
$\tphmed=-3440$\,kyr (90\% CI $-3630$ to $-3260$\,kyr),
$\dphmed=20.3$\,pc (90\% CI 18.0--22.9\,pc).

\object{Hip 71683} (\object{$\alpha$ Centauri A}). This is not in \gdr{2} on account of its brightness.

\object{Hip 71681} (\object{$\alpha$ Centauri B}). This is not in \gdr{2} on account of its brightness.

\object{Hip 70890} (\object{Proxima Centauri}).  This is \object{\gdr{2}  5853498713160606720}.
It has no radial velocity in \gdr{2}, but the parallax and proper motion agree with Hipparcos to within 2$\sigma$. Given their high precision, the perihelion parameters using the new astrometry together with the old radial velocity are consistent with the result in paper 1 (now more precise).

\object{Hip 3829}  (\object{van Maanen's star}). This is \object{\gdr{2}  2552928187080872832}. 
It has no radial velocity in \gdr{2}, but the parallax and proper motion agree with Hipparcos to better than 1$\sigma$. Given their high precision, the perihelion parameters using the new astrometry with the old radial velocity are consistent with the result in paper 1 (now more precise).

\object{Hip 42525}. Two \gdr{2} sources with the same parallax, proper motion, and radial velocity as each other (within their uncertainties) match this. They are \object{\gdr{2}  913394034663258752} and \object{\gdr{2}  913394034663259008}. The proper motion and radial velocity agree with the data used in paper~1, but the \gdr{2} parallax of $6.1 \pm 0.8$\,mas is quite different from the Hipparcos-2 value of $68.5 \pm 15.5$\,mas adopted in paper~1.
It was noted in that paper, however, that the Hipparcos-2 parallax was almost certainly corrupted by a close companion, and that the Hipparcos-1 solution of  $5.1 \pm 4.3$\,mas was more plausible. This value is in fact consistent with the new \gdr{2} one, which is five times more precise. It is (they are) no longer a close encounter.

\section{Completeness correction and the encounter rate}\label{sec:completeness}

\subsection{Principle of the completeness model}

\gdr{2} does not identify all encounters, primarily because of its faint-end magnitude limit for radial velocities. To compute an encounter rate over a specified time and distance window we must correct for this. 
This was done in paper 2 using a simple analytic model. Although insightful and easy to work with, that model made severe assumptions of isotropic spatial and homogeneous velocity distributions. 
Here we take the same conceptual approach to building a completeness map, but
replace the analytic model with a more realistic Galaxy simulation,
in which the spatial and kinematic distributions are also self-consistent.
We first simulate the positions and velocities of all stars in the nearby Galaxy (``mock full Galaxy''), and trace the orbits backwards and forwards in time (with LMA) to determine the distribution of encounters in perihelion coordinates. Call this
$\fmod(\tph, \dph)$,  the 
number of encounters per unit perihelion time and distance.
We then repeat this including only the stars which would have full 6D kinematic information in \gdr{2} (``mock \gaia-observed Galaxy''), then trace orbits to give $\fexp(\tph, \dph)$.
The ratio of these two distributions gives the completeness map, $C(\tph,\dph)$, which can be interpreted as the fraction of encounters at a specific perihelion time and distance that will actually be observed (in our sample).

\subsection{The model}\label{sec:themodel}

To generate our mock full Galaxy we use Galaxia \citep{2011ApJ...730....3S} to sample stars following a Besancon-like \citep{2003A&A...409..523R} model
distribution of the Galaxy, in a similar way to how we generated the mock \gdr{2} catalogue described in \cite{2018arXiv180401427R}. The main difference is that we now apply no magnitude cut, but
instead a distance cut, to keep the computations manageable.
We use 3\,kpc, as only stars with radial velocities above 200\,\kms\ would reach the Sun within 15\,Myr.
(Our final completeness-corrected encounter rate only uses the model out to 5\,Myr, so only mock stars beyond 3\,kpc with radial velocities above 600\,\kms\ will be neglected.)
We checked on a subsample of the data that increasing this limit to 10\,kpc made negligible difference to the distribution of encounters.  
We then computed the perihelion parameters of all stars using the LMA. 
Although the LMA is not an accurate model for long paths in the Galaxy, the inaccuracy introduced by this for the distribution as a whole will be small. The impact is further diminished given that we are ultimately interested only in the {\em ratio} of the two encounter distributions from these two models.
 
\begin{figure}
\begin{center}
\includegraphics[width=0.5\textwidth, angle=0]{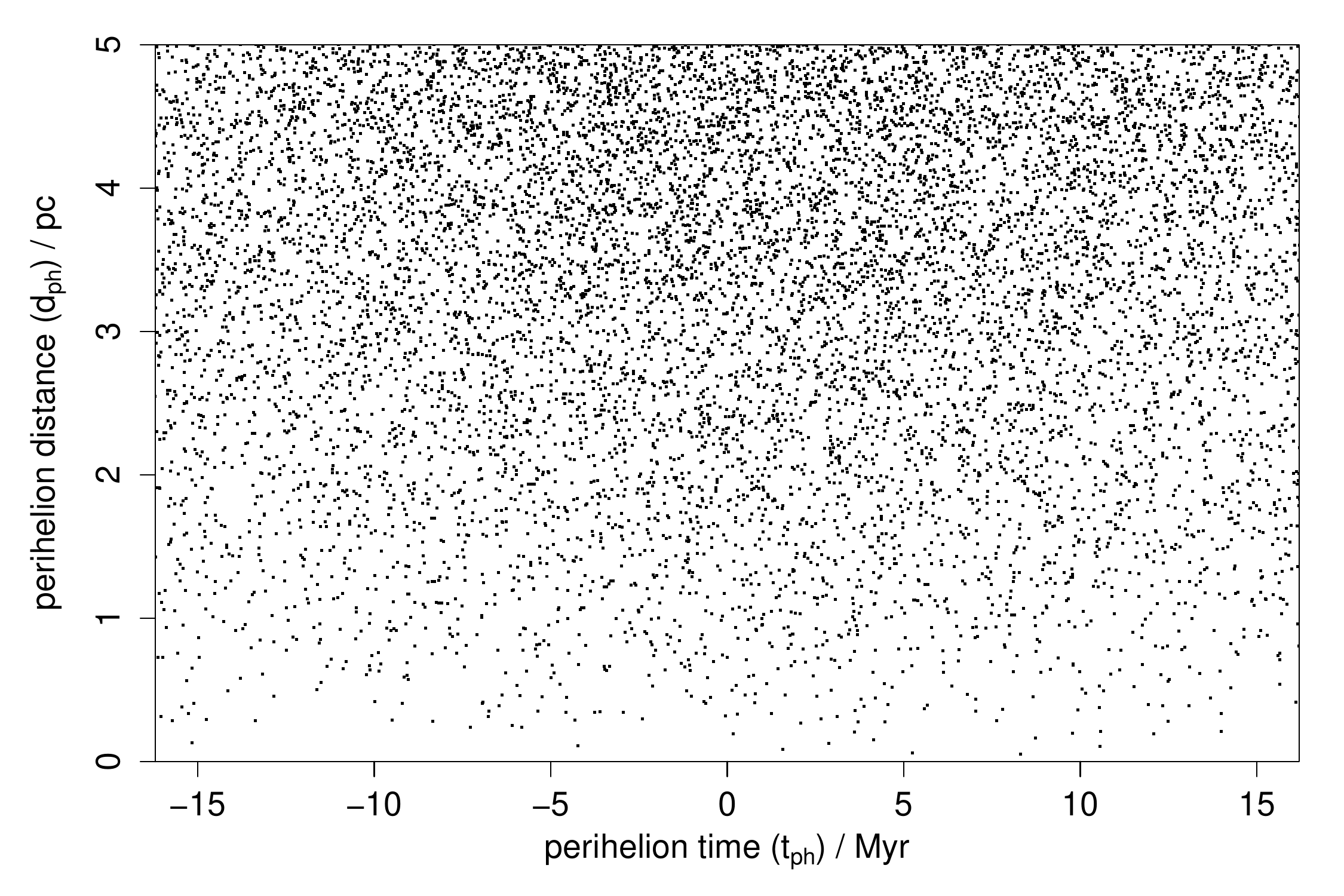}
\caption{Perihelion times and distances for star samples from the mock full Galaxy (no \gaia\ selection). Each star is represented by a single sample.
\label{fig:completeness_allenc_samples}}
\end{center}
\end{figure}

\begin{figure}
\begin{center}
\includegraphics[width=0.5\textwidth, angle=0]{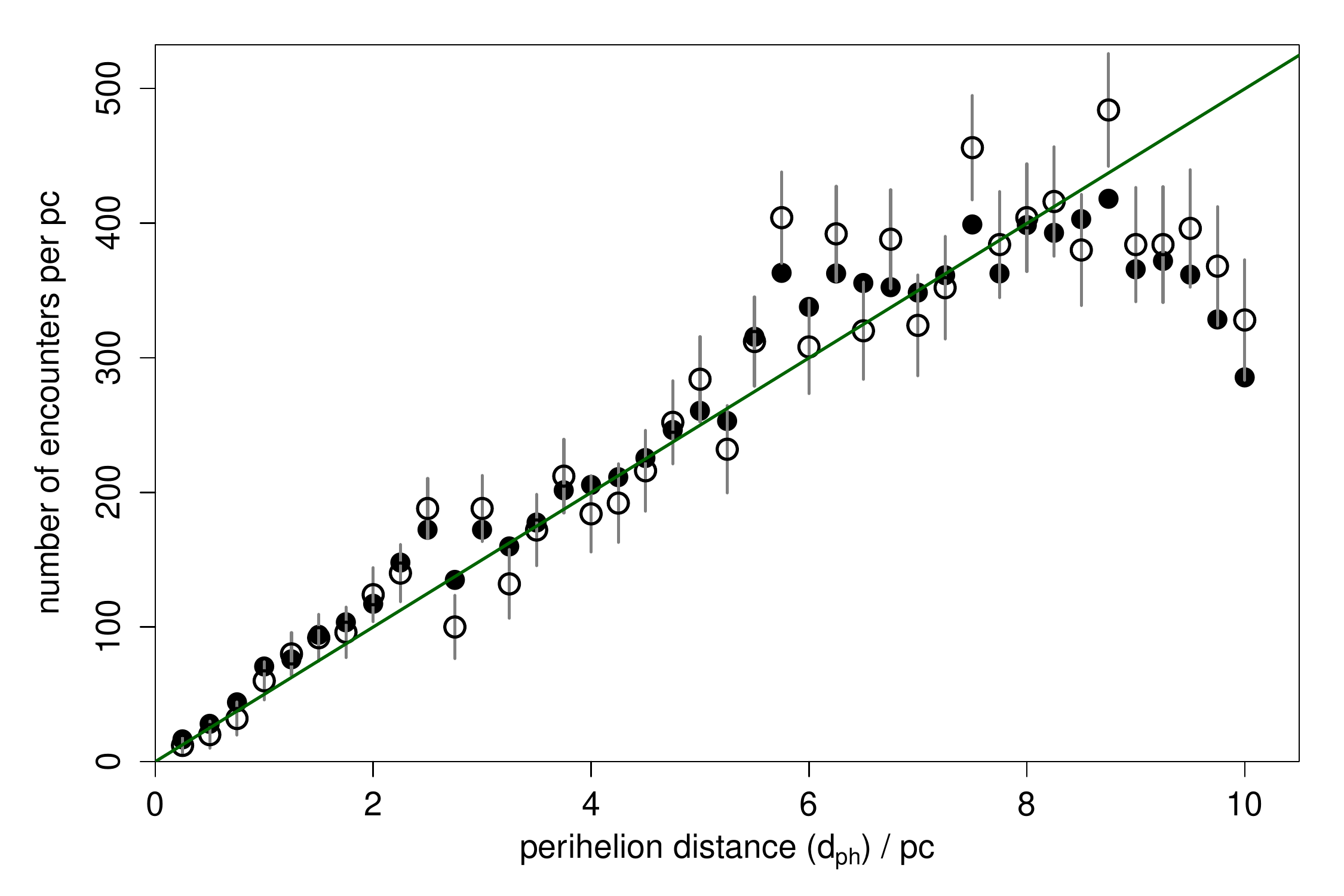}
\caption{The variation of the number of observed encounters per unit perihelion distance as a function of perihelion distance (for the filtered sample for all $\tph$).
The open circles count the encountering stars discretely, using $\dphmed$ as the distance estimate for each star.
The filled circles count each surrogate for each encountering star separately. The green line is the distribution we expect
under the assumption that the distribution is linear.
It has a gradient of 50.0\,pc$^{-1}$.
(The number of encounters within some distance is the integral of this, i.e.\ $\propto \dph^2$.)
The data do not follow this near to the limit (at 10\,pc) because not all stars found in the LMA-based selection actually come within 10\,pc after doing the orbital integration.
The error bars show the Poisson noise computed from the theoretical relation (only attached to the open points to avoid crowding).
\label{fig:fobs_dph}}
\end{center}
\end{figure}

Figure~\ref{fig:completeness_allenc_samples} shows the positions in perihelion parameter space of individual stars from the sampled mock full Galaxy.  The density of encounters shows essentially no dependence on $\tph$ over the range shown. This is different from what was found from the model in paper 2 which showed a drop off to large $|\tph|$. This was in part due to the spatial distribution adopted (and its inconsistency with the velocity distribution adopted). In the present model we also do not see the drop in density towards $\tph=0$. The encounter density in Figure~\ref{fig:completeness_allenc_samples} does show a strong dependency on $\dph$. Closer inspection shows that the number of encounters per unit distance varies linearly with $\dph$, i.e.\ the number of encounters within $\dph$ varies as $\dph^2$.  This is what the simple model in paper 2 predicts (see section 4.2 of that paper, which also explains why it does not vary as $\dph^3$, as one may initially expect).  To construct the completeness map we therefore replace the 2D distribution in Figure~\ref{fig:completeness_allenc_samples} with a 1D distribution $\fexp(\tph, \dph) = a \, \dph$, where the constant $a$ is fit from the simulated data.  The distribution of the real encounters also shows this linear variation out to several pc, as shown in Figure~\ref{fig:fobs_dph}.

To produce the mock \gaia-observed Galaxy we query the mock catalogue of \cite{2018arXiv180401427R} using
the following ADQL query

{\tiny 
\begin{verbatim}
SELECT parallax, pmra, pmdec, radial_velocity
FROM gdr2mock.main
WHERE phot_g_mean_mag <= 12.5 
AND teff_val > 3550 AND teff_val < 6900
\end{verbatim}
}

\noindent The \teff\ limit simulates the limit on the radial velocities published in \gdr{2}
\citep{2018arXiv180409372K}, and the magnitude limit is the selection we will apply to the observed encounter sample to compute the completeness-corrected encounter rate below.
We do not specifically account for the incompleteness at the bright end in \gdr{2} because it is poorly defined. There are so few bright stars in the model that this is anyway a minor source of error.
This above query delivers 4.4 million stars. 
(As a comparison, the number of stars in \gdr{2} with complete 6D kinematic data brighter than $\gmag=12.5$\,mag is 3.6 million.)

Given the nonlinear relationship between the 6D kinematic data and the perihelion parameters for a star, symmetric (Gaussian) errors in the former can lead to asymmetric distributions in the latter. In other words, measurement errors in the real Gaia observations can lead to stars being preferentially scattered into or out of some part of the perihelion parameter space. We accommodate this in our model by replacing each star in the mock Gaia-observed Galaxy with a set of 100 noise-perturbed samples generated using a simple error model. For this we use a 4D-Gaussian PDF (the \gaia-uncertainties on RA and Declination are negligible), with the following standard deviations, obtained from inspection of the \gdr{2} uncertainties shown in \cite{2018arXiv180409366L}:
$\sigma(\parallax)=0.068$\,mas; $\sigma(\pmra)=0.059$\,\maspyr;  $\sigma(\pmdec)=0.041$\,\maspyr;
$\sigma(\vr)=0.8$\,\kms. We neglect correlations as there is no reliable model for these.
Just as was done with the surrogates for the real data, the orbits of each of the samples are traced (but here using LMA rather than a potential).

\begin{figure}
\begin{center}
\includegraphics[width=0.5\textwidth, angle=0]{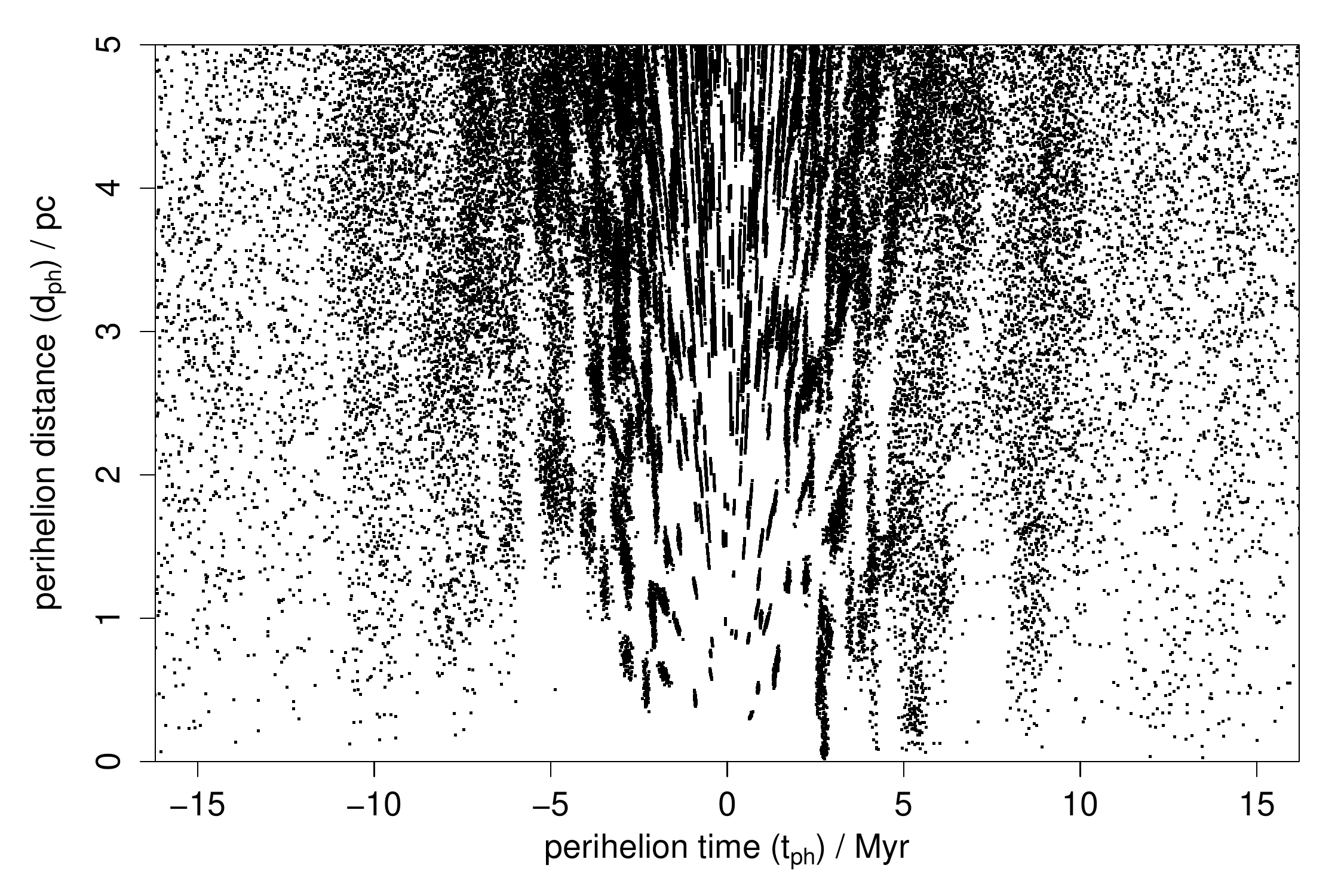}
\caption{Perihelion times and distances for star samples from the mock \gaia-observed Galaxy. Each star is represented by 100 samples. This is just a window on a much larger set of encounters, so includes samples of stars which 
are predominantly outside the range shown (e.g.\ the median perihelion position is outside).
\label{fig:completeness_obsenc_samples}}
\end{center}
\end{figure}

Figure~\ref{fig:completeness_obsenc_samples} shows the resulting positions in perihelion parameter space of the samples from the mock \gaia-observed Galaxy.  In principle this is a model for the observations shown in Figure~\ref{fig:dph_vs_tph_0to5pc_samples}. They are not the same due to shot noise, imperfect modelling of the (complex) \gdr{2} selection function, and in particular the fact that Galaxia is not an exact model of our Galaxy.  A particular difference is that the mock catalogue shows more encounters at larger perihelion times. There are several possible causes for this. One possibility is differing velocity distributions: if stars in the mock catalogue were generally slower, then the most distant visible stars would arrive at larger perihelion times.  An identical reproduction of the observational distribution is not necessary, however, because the completeness is dependent primarily on the change from the mock full Galaxy to the mock \gaia-observed Galaxy.

As an aside, we can use the mock \gaia-observed simulation to try to explain features in the observed perihelion distribution (e.g.\ Figure~\ref{fig:dph_vs_tph_0to5pc_witherrors}). In particular, the gap at small $|\tph|$ might be explained by the \teff\ cut. This cut removes cool M dwarfs, which are also
intrinsically faint. To be observable now, they must therefore be relatively near, in which case they will all encounter relatively soon (past or future). M dwarfs are highly abundant, so would dominate the encounters at small $|\tph|$. Yet they have been removed by the \teff\ cut. A similar argument in principle also applies to the brightest stars -- many of which are bright because they are close by, and thus near to encounter now -- which are also missing in \gdr{2}, but these are fewer, so their net contribution is smaller.

\begin{figure}
\begin{center}
\includegraphics[width=0.5\textwidth, angle=0]{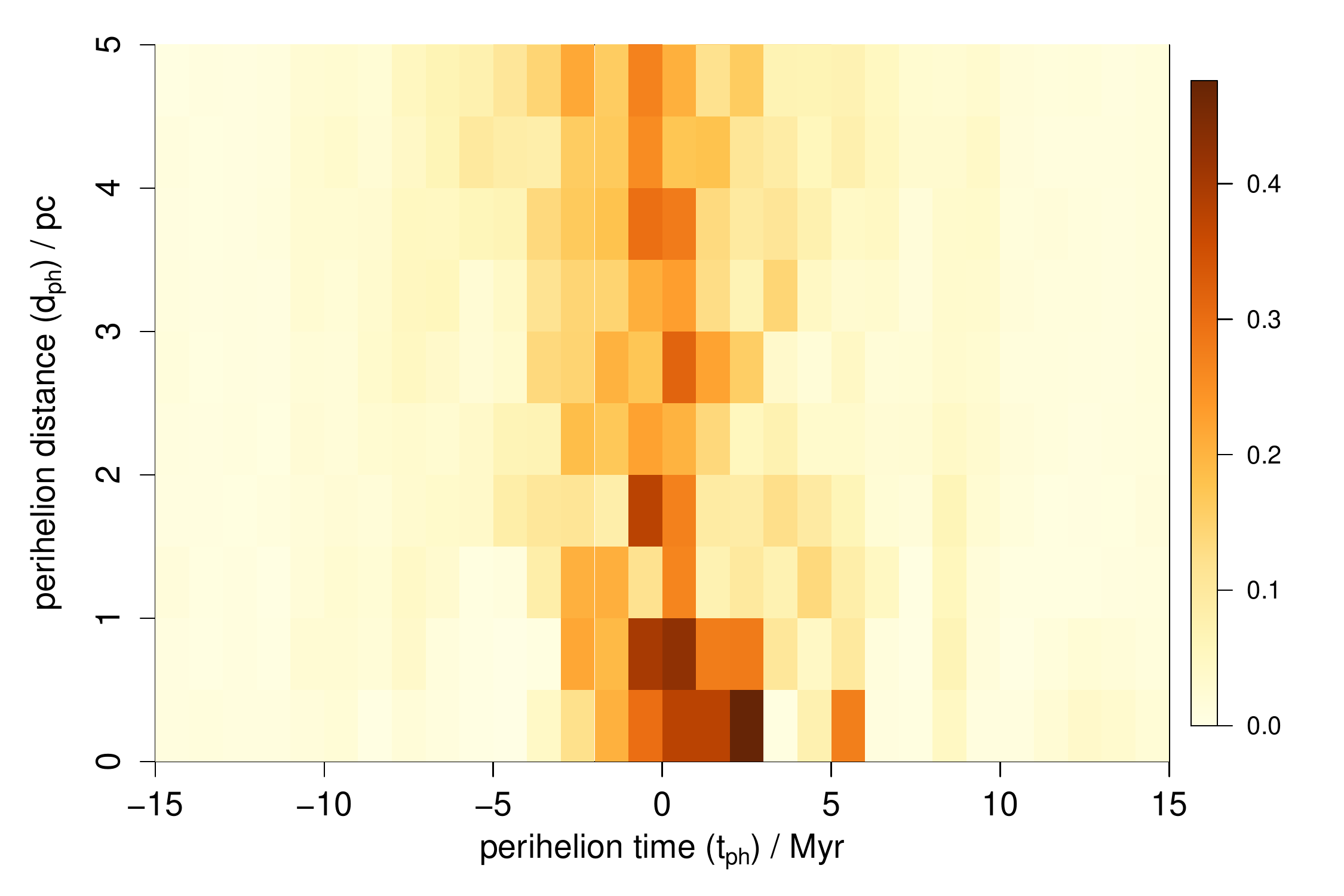}
\caption{The completeness map, $C(\tph, \dph)$, which can be interpreted as the probability of observing an encounter in the selected \gdr{2} sample as a function of its perihelion time and distance. ``Selected'' here means for $\gmag<12.5$\,mag in the filtered sample.
\label{fig:completeness_map}}
\end{center}
\end{figure}

We can now build the completeness map. We bin the distribution in Figure~\ref{fig:completeness_obsenc_samples}, and divide by the number of samples, to get a binned distribution of the (fractional) number of encountering stars per bin. We then divide this by the expected (continuous) distribution from the mock full Galaxy, which was just $\fexp(\tph, \dph)  = a \, \dph$, to give our completeness map.
This is shown in Figure~\ref{fig:completeness_map}. 
The map was computed and fitted over the range $-15$ to $+15$\,Myr and 0 to 10\,pc to improve the fit to $\fexp(\tph, \dph)$ (by reducing the shot noise), but is only shown over a smaller $\dph$ range (and will only be used over a narrower $\tph$ range too).
The completeness ranges between nearly zero and 0.48, with an average value (over the bins) of 0.09 for $|\tph|<10$\,Myr and 0.14 for $|\tph|<5$\,Myr.
The binning of Figure~\ref{fig:completeness_obsenc_samples} resulted in some cells with no encounters, which would lead to zero completeness (and thus a divide-by-zero when we come to use it). We overcame this by replacing zero-valued cells with the mean of their neighbours. We experimented with producing a smoother map via kernel density estimation or fitting a smoothing spline, but the results were too sensitive to the fitting parameters and suffered from edge effects (which could be mitigated by expanding the grid, but at the expense of sensitivity to low completeness values at larger times).

\subsection{An estimate of the encounter rate}\label{sec:encounter_frequency}

If a star is observed to be encountering at position $\tph,\dph$, then the completeness-corrected (fractional) number of
encounters corresponding to this is simply $1/C(\tph,\dph)$. When each star is instead represented by a set of $N_{\rm sur}$ surrogates, the completenesses at which are $\{C_i\}$, the completeness-corrected number of encounters is
\begin{equation}
\ncor \,=\, \frac{1}{N_{\rm sur}}\sum_i\frac{1}{C_i} \ .
\label{eqn:ncor}
\end{equation}
This same expression applies when we have a set of stars represented by surrogates over some range (``window'') of $\tph$ and $\dph$ with corresponding completenesses $\{C_i\}$. $N_{\rm sur}$ is still the number of surrogates per stars. \ncor\ is then the completeness-corrected number of encounters in the window.

Random errors in \ncor\ come from two main sources: (i) Poisson noise from the finite number of encounters observed; (ii) noise in the completeness map. 
The first source is easily accommodated.
The fractional number of encounters observed in a window, \nenc, is just the sum of the fraction of all surrogates per star which lie in that window. The signal-to-noise ratio in this is $\sqrt{\nenc}$, so the standard deviation in \ncor\ due to this alone would be $\ncor/\sqrt{\nenc}$. 
This does not include any noise due arising from the original sample selection (e.g.\ from changes in the filtering).
The second source can be treated with a first order propagation of errors in equation~\ref{eqn:ncor}
\begin{equation}
\delta\ncor \,=\, \frac{1}{N_{\rm sur}}\sum_i\frac{1}{C_i}\frac{\delta C_i}{C_i} \ .
\label{eqn:ncor_comperr}
\end{equation}
In practice it is difficult to determine $\delta C_i$. We have experimented with varying the approach to constructing the completeness map. We approximate the uncertainties arising thereby as a constant fractional uncertainty of $f_c = \delta C_i/C_i = 0.1$. 
In that case equation \ref{eqn:ncor_comperr} can be written $\delta\ncor = f_c\ncor$. Combining this with the Poisson term for (i) we may approximate the total random uncertainty in \ncor\ as
\begin{equation}
\sigma(\ncor) \,=\, \ncor\left(\frac{1}{\nenc} + f_c^2\right)^{1/2} \  .
\label{eqn:ncor_sigma}
\end{equation}

To compute \nenc\ we use just the filtered encounter sample with $\gmag<12.5$\,mag -- 2522 stars -- as this is what was also used to build the completeness model. We do not remove bogus encounters, as their number is small. (Identifying them is a laborious, manual process that we have only done for the 31 encounters out to 1\,pc.)
Within the window $|\tph|<5$\,Myr, $\dph<5$\,pc we have \nenc\,=\,463.4 (the sum in equation ~\ref{eqn:ncor} is over 926\,736 surrogates), which is an {\em uncorrected} rate of 46 encounters per Myr within 5\,pc.
This compares to 639 stars with $\tphmed$ and $\dphmed$ in this range (i.e.\ neglecting the uncertainties gives an 
 {\em uncorrected} rate of 64 stars per Myr within 5\,pc.)
Applying the completeness correction as outlined above yields $\ncor = 4914 \pm 542$ encounters,
which corresponds to $491 \pm 54$ encounters per Myr within 5\,pc. 
We could calculate the corrected rates for smaller upper limits on $\dph$, but such results are sensitive to the use of fewer bins in the completeness map and are more affected by the scattering of the surrogates.
We instead scale the value found for 5\,pc using the expectation that the number of encounters within some distance grows quadratically with distance. 
(Figure~\ref{fig:fobs_dph} shows that out to 5\,pc the scaling is as expected, unaffected by the drop-off near to 10\,pc.)
This gives encounter rates of $78.6 \pm 8.7$ per Myr within 2\,pc, and $19.7 \pm 2.2$ per Myr within 1\,pc. We adopt these as our final encounter rates.

In \citetalias{2018A&A...609A...8B}, using TGAS and an analytic completeness correction, we derived an encounter rate within 5\,pc of $545 \pm 59$ per Myr (which scales to $21.8 \pm 2.4$ per Myr within 1\,pc). This is consistent with what we find here.  \cite{2001A&A...379..634G}  obtained a rate of $11.7 \pm 1.3$ per Myr within 1\,pc 
using Hipparcos. Possible reasons for this discrepancy are discussed in \citepalias{2018A&A...609A...8B}.

Choosing a smaller time window, $|\tph|<2.5$\,Myr and $\dph<5$\,pc, we have \nenc\,=\,319.4 and a completeness-corrected encounter rate of
$373 \pm  44$ per Myr within 5\,pc. This is 1.5$\sigma$ times smaller than that obtained with the larger time window. This may suggest some time variability in the encounter rate, although this is hard to distinguish given the difficulty of propagating all the uncertainties.
It is also clear that equation~\ref{eqn:ncor} is rather sensitive to small values of the completeness. 
Over the window $|\tph|<5$\,Myr and $\dph<5$\,pc, 3462 of the 926736 surrogates (0.4\%) have $C_i<0.01$. It is an unfortunate and unavoidable fact that it is the low completeness regions which contribute the largest uncertainty to any attempt to correct the encounter rate.

\section{Conclusions}\label{sec:conclusions}

We have identified the closest stellar encounters to the Sun from among the 7.2 million stars in \gdr{2} that have 6D phase space information. Encounters were found by integrating their orbits in a smooth gravitational potential. The correlated uncertainties were accounted for by a Monte Carlo resampling of the 6D likelihood distribution of the data and integrating a swarm of surrogate particles. The resulting distributions over perihelion time, distance, and speed are generally asymmetric, and are summarized by their 5th, 50th, and 95th percentiles.

We find 31, 8, and 3 stars which have come -- or which will come -- within 1\,pc, 0.5\,pc, and 0.25\,pc of the Sun, respectively.  These numbers drop to 26, 7, and 3 when we remove likely incorrect results (``bogus encounters'') following a subjective analysis (including visual inspection of images). 
Quality metrics in the \gaia\ catalogue are not calibrated and are hard to use to identify sources with reliable data. Thus some of the stars in our encounter list are sure to be bogus, and we are sure to have missed some others for the same reason. In particular, a number of encounters with unexpected positions in the CMD have large values of the astrometric unit weight error. These are not necessarily poor astrometric solutions. At least some could be main sequence--white dwarf binaries.

The closest encounter found is \object{Gl 710}, long known to be a close encounter, now found to come slightly closer and with slightly better determined perihelion parameters. 
Most of the other encounters found are discovered here, including 25 within 1\,pc.
Using newly available masses for 98\% of the sample computed by Fouesneau et al.\ (in preparation) from \gaia\ astrometry and multiband photometry, we compute the impulse transfer to the Oort cloud using the impulse approximation. For both the $1/\dph$ or $1/\dph^2$ dependencies, \object{Gl 710} induces the largest impulse.  \cite{2016A&A...595L..10B} studied the impact of this star on the Oort cloud in some detail using the encounter parameters from the first \gaia\ data release. They found that it would inject a large flux of Oort cloud comets toward the inner solar system. Given that the encounter parameters are not significantly changed in \gdr{2}, this conclusion still holds. It remains to be studied what the cumulative effect is of the many more encounters found in our study.

The main factors limiting the accuracy of our resulting perihelion parameters are: for most stars, the accuracy of the radial velocities; for distant stars, the accuracy of our potential model; 
and for some stars, the neglect of possible binarity.

Our sample is not complete. The main limitation is the availability of radial velocities in \gdr{2}.
99\% of the stars (whether in the unfiltered or filtered sample) are brighter than $\gmag=13.8$\,mag. Published radial velocities are also limited to stars in the approximate \teff\ range 3550--6900\,K, thereby limiting the number of
(numerous) late-type stars as well as massive stars.
Both \teff\ and \gmag\ should be extended in subsequent \gaia\ data releases. Correcting for this incompleteness using a model based on the Galaxia simulation, we infer the encounter rate averaged over the past/future 5\,Myr to be $491 \pm 54$\,Myr$^{-1}$ within 5\,pc. When scaled to encounters within 1\,pc, this is $19.7 \pm 2.2$\,Myr$^{-1}$. We caution, however, that the accuracy of this rate is limited by the completeness model assumptions and fitting the resulting completeness map, as well as the distribution of the actual encounters. It may also be overestimated if there is a large fraction of sources with spuriously large parallax values that are not accounted for by their formal uncertainties.

The other source of incompleteness is missing bright stars in \gdr{2}. Bright stars saturate even with the shortest CCD gate on \gaia, making them harder to calibrate. Astrometry for these will only be provided in later data releases. Although few in number (and therefore acceptably neglected by our completeness correction), some bright stars are currently nearby and/or massive, so encounter parameters more accurate than those obtained with Hipparcos could ultimately reveal some important encounters. The case of \object{gamma Microscopii} -- no longer a close encounter in \gdr{2} -- has already been discussed.  
Another case is the A2 dwarf \object{zeta Leporis}, computed in \citetalias{2015A&A...575A..35B} to have $\dphmed=1.30$\,pc 850\,kyr ago. 
It is in \gdr{2} but without a radial velocity. Using the old radial velocity we find its perihelion parameters to be rather similar
($\tphmed=-860$\,kyr, $\dphmed=1.43$\,pc).  
\object{Sirius}, \object{Altair}, and \object{Algol} are not in \gdr{2}.

There are no doubt many more close -- and probably closer -- encounters to be discovered in future \gaia\ data releases.

\begin{acknowledgements}

This work is based on data from the European Space Agency (ESA) mission \gaia\ (\url{https://www.cosmos.esa.int/gaia}), processed by the \gaia\ Data Processing and Analysis Consortium (DPAC, \url{https://www.cosmos.esa.int/web/gaia/dpac/consortium}). Funding for the DPAC has been provided by national institutions, in particular the institutions participating in the \gaia\ Multilateral Agreement.
We thank Lennart Lindegren for discussions about the \gaia\ astrometric solution and Ted von Hippel for discussions about white dwarfs.
This work was funded in part by the DLR (German space agency) via grant 50 QG 1403. We are grateful for the
availability of the Simbad object database and Aladin sky atlas, both developed and provided by CDS, Strasbourg.

\end{acknowledgements}

\bibliographystyle{aa}
\bibliography{stellar_encounters,gaia}

\appendix

\section{Gaia archive query}\label{appendix:query}

Below is the ADQL query used to select stars from \gdr{2} which have perihelion distances less than 10\,pc according to the linear motion approximation (specified by equation 4 in \citetalias{2015A&A...575A..35B}).
\tiny{
\begin{verbatim}
SELECT
source_id, ra, dec, parallax, pmra, pmdec, 
radial_velocity, ra_error, dec_error, parallax_error, 
pmra_error, pmdec_error, radial_velocity_error,
ra_dec_corr, ra_parallax_corr, ra_pmra_corr, 
ra_pmdec_corr, dec_parallax_corr, dec_pmra_corr, 
dec_pmdec_corr, parallax_pmra_corr, parallax_pmdec_corr, 
pmra_pmdec_corr, visibility_periods_used, 
phot_g_mean_mag, bp_rp, astrometric_chi2_al, 
astrometric_n_good_obs_al, astrometric_excess_noise, 
astrometric_excess_noise_sig, rv_nb_transits, l, b
FROM gaiadr2.gaia_source
WHERE (parallax IS NOT NULL) AND (parallax > -0.029) 
AND (radial_velocity IS NOT NULL) AND 
( 
  ( 
    ( 1000*4.74047*sqrt(power(pmra,2)+power(pmdec,2))
      / power(parallax+0.029,2) 
    ) 
    / sqrt( (power(pmra,2)+power(pmdec,2)) 
              * power(4.74047/(parallax+0.029),2)
              + power(radial_velocity,2) 
          ) 
  ) < 10
)
\end{verbatim}
}

\end{document}